\documentclass[journal]{IEEEtran}
\usepackage[utf8]{inputenc}
\usepackage{amsthm,authblk,color,booktabs}
\usepackage{graphicx,epstopdf}
\usepackage{algorithm,algcompatible}
\usepackage{epsfig}
\usepackage{amsfonts}
\usepackage{bbold}
\usepackage{cite}
\usepackage{multirow}
\usepackage{amsmath}
\usepackage{neuralnetwork}
\usepackage{url}
\usetikzlibrary{shapes,arrows,positioning,shadows,trees,calc}
\usetikzlibrary{arrows.meta}

\title{A General Method for Calibrating Stochastic Radio Channel Models with Kernels}

\author{Ayush Bharti, Fran\c{c}ois-Xavier Briol, Troels Pedersen 
\thanks{Ayush Bharti and Troels Pedersen are with the Department of Electronic Systems, Aalborg University, 9220 Aalborg East, Denmark (e-mail: \{ayb, troels\}@es.aau.dk). 

Fran\c{c}ois-Xavier Briol is affiliated with the Department of Statistical Science at University College London and the Data-Centric Engineering Programme at The Alan Turing Institute, London, United Kingdom (email: f.briol@ucl.ac.uk).

This work is supported by the Danish Council for Independent Research, grant no. DFF 7017-00265. Fran\c{c}ois-Xavier Briol was supported by the Lloyds Register Foundation Programme on Data-Centric Engineering at The Alan Turing Institute under the EPSRC grant [EP/N510129/1].
}}

\begin{document}

\maketitle

\newcommand{\btheta}{\boldsymbol{\theta}}
\newcommand{\tmax}{t_{\mathrm{max}}}
\newcommand{\nreal}{N_{\mathrm{obs}}}
\newcommand{\nsim}{N_{\mathrm{sim}}}
\newcommand{\ysim}{\mathbf{Y}_{\mathrm{sim}}}
\newcommand{\sobs}{\mathbf{s}_{\mathrm{obs}}}

\begin{abstract}
    Calibrating stochastic radio channel models to new measurement data is challenging when the likelihood function is intractable. The standard approach to this problem involves sophisticated algorithms for extraction and clustering of multipath components, following which, point estimates of the model parameters can be obtained using specialized estimators. We propose a likelihood-free calibration method using approximate Bayesian computation. The method is based on the maximum mean discrepancy, which is a notion of distance between probability distributions. Our method not only by-passes the need to implement any high-resolution or clustering algorithm, but is also automatic in that it does not require any additional input or manual pre-processing from the user. It also has the advantage of returning an entire posterior distribution on the value of the parameters, rather than a simple point estimate. We evaluate the performance of the proposed method by fitting two different stochastic channel models, namely the Saleh-Valenzuela model and the propagation graph model, to both simulated and measured data. The proposed method is able to estimate the parameters of both the models accurately in simulations, as well as when applied to 60 GHz indoor measurement data. 
\end{abstract}

\vskip0.5\baselineskip
\section{Introduction}

Stochastic channel models are used to simulate the behavior of the radio channel in order to test the performance of communication and localization systems. Often models are flexible enough to be applied to different scenarios, provided that their parameters can be adjusted accordingly. Adjustment of the model parameters based on data collected from measurement campaigns is called \textit{calibration} (or inference). Calibration is usually challenging since most state-of-the-art stochastic radio channel models have intractable likelihood functions. This renders usual inference techniques such as maximum likelihood estimation or standard Bayesian inference inapplicable.

\begin{figure*}
	 \centering
		 \includegraphics[trim={30 390 580 20}, clip,  width = 0.75\textwidth]{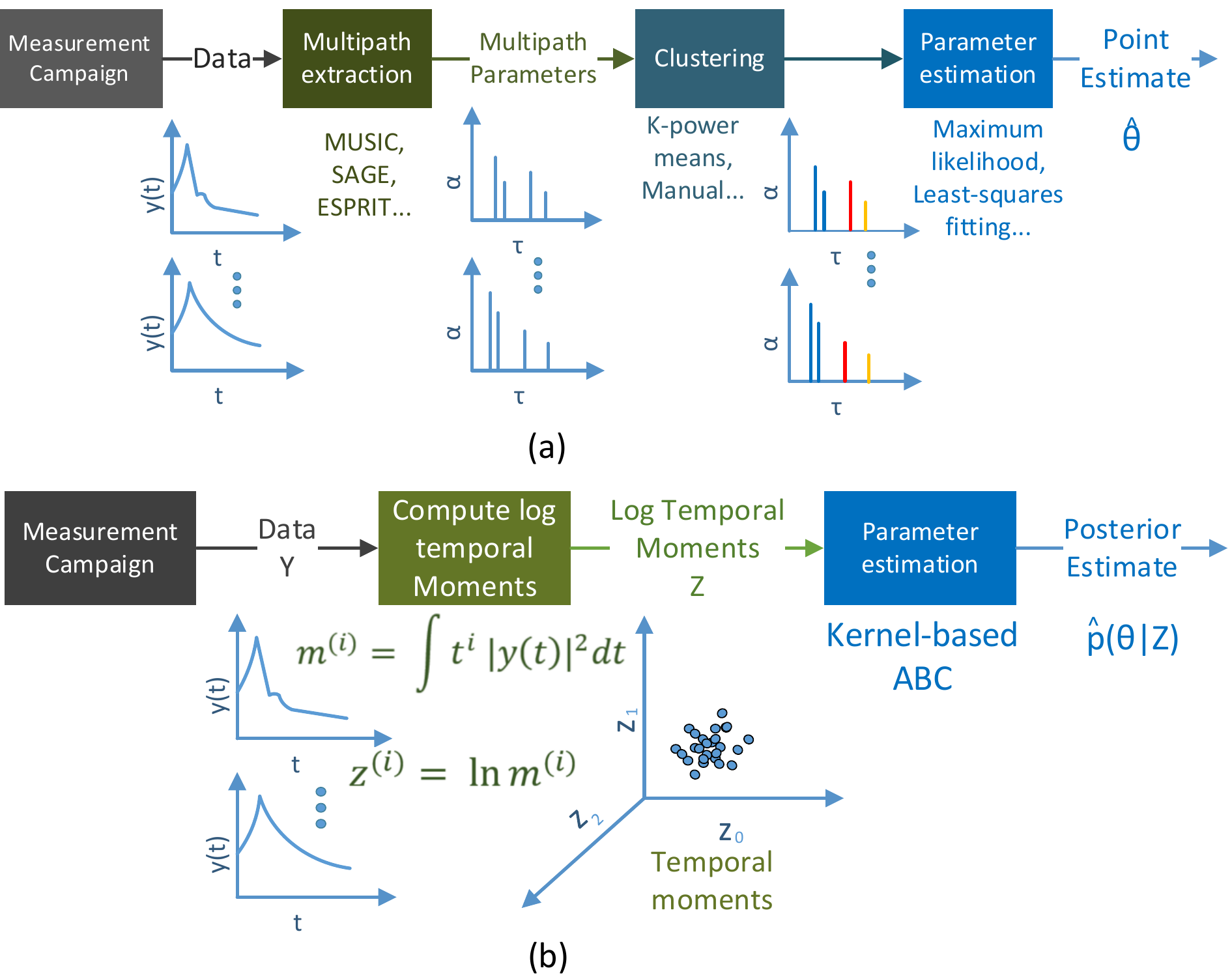}
	 \caption{Methodologies for calibration of stochastic radio channel models: (a) State-of-the-art methodology based on multipath extraction and clustering; (b) proposed method based on generic summaries (here exemplified by log-temporal moments) and approximate Bayesian computation (ABC).} 
	 \label{fig:methodology} 
 \end{figure*}

Instead of solving the whole calibration problem at once, it is wide-spread practice (e.g. \cite{Turin1972, Saleh1987, Haneda, METIS, WINNER, Gustafson, COST, Li2019, Yang2020}) to split the task into intermediate steps as outlined in Fig.~\ref{fig:methodology}(a). The first step involves resolving the multipath components, i.e. estimating path parameters including delays, directions, and complex gains. This task can be carried out using high-resolution algorithms such as MUSIC, SAGE, and RiMAX, among others, see e.g \cite[Ch. 5]{Yin2018} for an overview. The second step is clustering of the extracted multipath components in the case of cluster-based models.  Clustering is either performed manually, as in \cite{Saleh1987}, or using automated algorithms such as \cite{Czink, Gentile, Ruisi}. In a final step, the model parameters are estimated from the extracted and clustered multipath components. 

Despite being widely applied, the multi-step approach  suffers from a range of issues, owing  to the composite nature of the methodology.  In particular, high-resolution and clustering methods, although very useful in analyzing and understanding the radio channel, are problematic when it comes to model calibration. These methods require implementation of sophisticated and specialized algorithms at each step, which involves a number of heuristic choices and settings which might be conflicting. An emblematic example is the assumption of  ``well separated" paths while extracting multipath components. The high-resolution methods are prone to estimation artifacts, especially if paths are not ``well separated". However, this conflicts with the inherent assumption in the clustering step that  multipaths arrive ``close" to each other. Consequently, even though the performance of high-resolution and clustering algorithms are thoroughly investigated in isolation, the accuracy of the applied multi-step calibration techniques is unknown. Moreover, the calibration technique needs to be tailored to the particular model at hand. While attempting to calibrate and compare different ultra-wideband models using a large database, Greenstein et al. in \cite{Greenstein2007} noted that \emph{``the problem in doing so is that there is no simple, clear and established method for extracting cluster model parameters from measured data"}. As a result, they were unable to fit the cluster model to their calibration data.

Calibration methods that by-pass the need to resolve the multipath components have been recently proposed. They have been used to calibrate the Turin model \cite{Turin1972}, the Saleh-Valenzuela (S-V) model \cite{Saleh1987} and the polarized propagation graph (PG) model \cite{RamoniTAP2019}. These calibration methods rely either on a Monte Carlo approximation of the likelihood \cite{Hirsch2020, AyushURSI}, the method of moments \cite{Wu2008, AyushSPAWC19}, or a summary-based likelihood-free inference framework \cite{AyushABC, RamoniMLAWPL, Bharti2020,AyushEuCAP21} such as approximate Bayesian computation (ABC). First developed in the field of population genetics in 1997, ABC has since become a popular method for calibrating models with intractable likelihoods in various fields, see \cite{Sisson2018} for an overview. The main drawback of the calibration methods \cite{Wu2008, AyushURSI, AyushSPAWC19} is their reliance on equations that explicitly link the moments of the summaries with the model parameters, or in case of \cite{Hirsch2020}, on the model-specific point process. These methods should therefore be re-derived for each new model. We encounter this to be a non-trivial task, and it may not even be possible for the more elaborate channel models. Similar problems exist in \cite{AyushABC, RamoniMLAWPL, Bharti2020} where a low-dimensional vector of statistics should be redesigned or trained using an autoencoder \cite{AyushEuCAP21} for the channel model at hand, which is not always trivial and may not generalize to other models. Moreover, summarizing the data leads to information loss that can hamper the accuracy of the parameter estimates. 

The aim of the present contribution is to propose a general method which can be applied to stochastic channel models of very different mathematical structure. This will be done without the need for specializing summaries, or extraction and clustering of multipaths. To achieve this, we follow the proposed calibration methodology depicted in Fig.~\ref{fig:methodology}(b). First, we map the channel measurements into easily computable log temporal moments. These moments are then used for calibration in an ABC framework, where we use the maximum mean discrepancy (MMD) \cite{Gretton2012JMLR} to compare  the distribution of simulated and measured data. The MMD has previously been used for frequentist inference in \cite{Briol2019MMD,Cherief-Abdellatif2019finite}, and in a Bayesian sense in \cite{Cherief-Abdellatif2019MMDBayes}. Specific ABC methods using kernels include \cite{Nakagome2013,Park2015,Mitrovic2016,Kisamori2020}, and the MMD has also been used to train generative adversarial networks in \cite{Dziugaite2015,Sutherland2017,Li2015GMMN}. These papers have shown MMD to be a powerful way to represent either data-sets or distributions, and as a result calibrate complex models. They have acted as inspiration for our work, but our algorithm specializes the approach to the problem of calibrating stochastic channel models. Our calibration method is automatic since it can be applied to different models without the need for further pre- or post- processing. Additionally, the method is able to account for model misspecification, which occurs when the model is not able to represent the data for any parameter setting. 

The rest of the paper is organized as follows. Section~\ref{sec:2} presents the model calibration problem. Section~\ref{sec:3} gives an overview of the MMD, and Section~\ref{sec:4} describes the proposed kernel-based ABC method. 
We  demonstrate the method's generality by calibrating the seminal S-V model, which is a clustered multipath model, and the propagation graph model, which is based on a different principle, using exactly the same data and procedure. Indeed, no other method able to do this is available in the open literature. In Section~\ref{sec:5}, the performance  is evaluated on simulated data and in Section~\ref{sec:6} on data from a 60~GHz indoor measurement campaign. We find that the S-V model is misspecified for the considered measurements, and hence fails to replicate its characteristics. Discussion and  concluding remarks are given in Sections ~\ref{sec:7} and~\ref{sec:8}, respectively.
\section{Stochastic Channel Model Calibration}\label{sec:2}

Consider the transfer function measurement of a linear, time-invariant radio channel in a single-input, single-output (SISO) setup using a vector network analyzer (VNA). The transfer function is measured at $N_s$ equidistant frequency points in the bandwidth $B$, resulting in a frequency separation of $\Delta f = B / (N_s - 1)$. The measured complex signal at the $n^\textup{th}$ frequency point, $Y_n$, is modeled as 
\begin{equation}
    \label{eq:receivedSignal}
    Y_n = H_n + W_n, \quad n = 0,1,\cdots, N_s-1,
\end{equation}
where $H_n$ is the transfer function sampled at the $n^\text{th}$ frequency and $W_n $ is the complex measurement noise. The additive noise samples are assumed independent and identically distributed (iid) at each frequency point, and are usually modeled as zero-mean circular symmetric complex Gaussian variables with variance $\sigma_W^2$. The time-domain signal, $y(t)$, is obtained by taking the discrete-frequency, continuous-time inverse Fourier transform of $Y_n$ as
\begin{equation}\label{eq:impulseResponse}
    y(t) = \frac{1}{N_s} \sum_{n=0}^{N_s-1} Y_n \exp(j2\pi n \Delta f t),
\end{equation}
periodic with a period of $\tmax = 1 / \Delta f$. Multiple realizations of the channel can be obtained by repeating the measurements $\nreal$ times, yielding an $\nreal \times N_s$ complex data matrix $\mathbf{Y}$. The data can be thought of as iid realizations from some unknown distribution, $\mathbb{Y}$, which is the true state of nature.

A stochastic model can be seen as a parametric family of distributions $\{\mathbb{P}_{\btheta}\}$ with a $p$-dimensional parameter vector $\btheta$ defined on some Euclidean space\footnote{The restriction to  parameters in $\mathbb R^p$ is only needed in the adjustment method described in Section~\ref{sec:adjustment}. The remaining part of the method can be used for more general parameter spaces, e.g. discrete, complex or subsets of $\mathbb R^p$. In this case, the adjustment algorithm should be modified to either accommodate or ignore such parameters.}. In the case of generative models such as the stochastic channel models, it is straightforward to simulate realizations of $\mathbf{Y}$ from the model, even though the distribution $\mathbb{P}_{\btheta}$ is unknown. Calibration then amounts to finding the $\btheta$ for which  the model output fits the observed data $\mathbf{Y}$ well, or in other words, to find the $\btheta$ such that $\mathbb{P}_{\btheta}$ is ``closest" to $\mathbb{Y}$. Standard calibration techniques involve the likelihood function of the model given $\mathbf{Y}$. For iid realizations, the likelihood function, denoted as $p(\mathbf{Y | \btheta})$, is the product of the probability density or mass function of $\mathbb{P}_{\btheta}$ evaluated at each of the data points in $\mathbf{Y}$.  

For most stochastic radio channel models, $p(\mathbf{Y | \btheta})$ is either intractable or cannot be approximated within reasonable computation time. Intractability here refers to the inability to numerically evaluate the likelihood function for a given value of $\btheta$. For intractable likelihood, the posterior, $p(\btheta | \mathbf{Y})$, also becomes intractable as it is  proportional to $p(\mathbf{Y | \btheta}) p(\btheta)$, where $p(\btheta)$ is the prior assumed on the parameters. An intractable likelihood prevents maximum likelihood estimation of $\btheta$ as well as  Bayesian inference via sampling of the posterior. This is the case for stochastic multipath models, such as the Turin and the S-V model, which were constructed with the ease of simulation in mind.

Since stochastic channel models are easy to simulate from given an arbitrary $\btheta$ value, likelihood-free inference is possible by comparing simulated data-sets to the observed data. Therefore, we need a method to compute distances between the data-sets which is challenging as the data-sets are high-dimensional, and may have possibly different sizes. We tackle this problem using distance metrics based on kernels, in particular the maximum mean discrepancy (MMD). 

\section{The Maximum Mean Discrepancy between Probability Distributions} \label{sec:3}

We now introduce the MMD which is a notion of distance between arbitrary probability distributions $\mathbb{P}$ and $\mathbb{Q}$ or data-sets. We aim to use MMD as a similarity measure within an ABC framework to compare simulated and observed data-sets. Note that we can identify any data-set $\{\mathbf{x}_1,\ldots,\mathbf{x}_n\}$ to an empirical distribution $\frac{1}{n} \sum_{i=1}^n \delta_{\mathbf{x}_i}$ where $\delta_{\mathbf x_i}$ denotes a distribution with mass one at $\mathbf x_i$ and $0$ otherwise. We restrict our discussion to distributions defined on $\mathbb{R}^d$. This section will provide further details on constructing the MMD \cite{Gretton2012JMLR,Muandet2017}.

\subsection{Kernels and The Maximum Mean Discrepancy (MMD)}

\begin{figure}
	 \centering
		 \includegraphics[trim={85 636 40 20}, clip,  width = 0.489\textwidth]{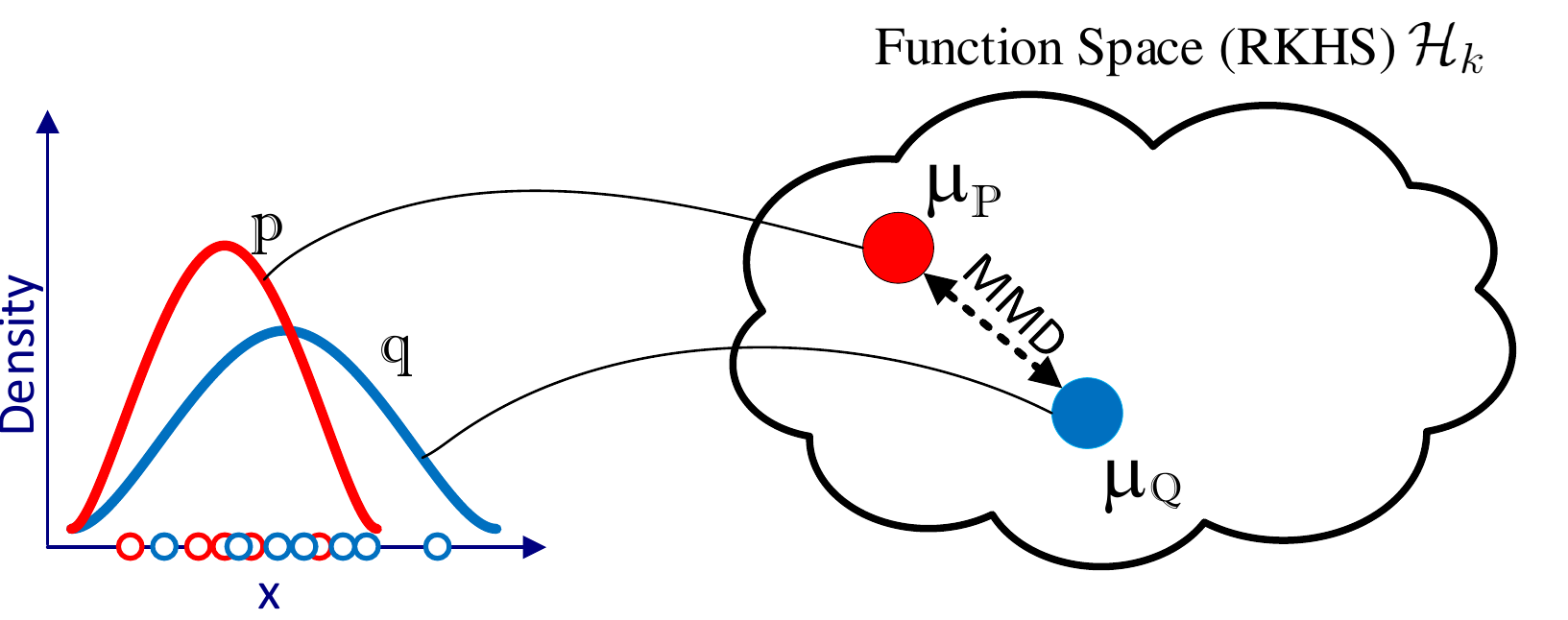}
	 \caption{Given a kernel $k$, the distributions $\mathbb{P}$ and $\mathbb{Q}$ are mapped to their kernel mean embeddings $\mu_{\mathbb{P}}$ and $\mu_{\mathbb{Q}}$ using Equation \ref{eq:meanEmbedding}. The MMD is obtained by computing the distance between $\mu_{\mathbb{P}}$ and $\mu_{\mathbb{Q}}$ in the function space $\mathcal{H}_k$, as expressed in Equation \ref{eq:MMD}. This figure is inspired by \cite{Muandet2017}.}
	 \label{fig:embedding}
 \end{figure}

The MMD consists of first mapping the distributions to a function space $\mathcal{H}_k$, then using the distance in that space to compare the mapped distributions. See  Fig.~\ref{fig:embedding} for an illustration. The mapping enables the use of distance defined on $\mathcal{H}_k$.

The spaces of functions to which we will map distributions are called reproducing kernel Hilbert space (RKHS). We denote the RKHS with $\mathcal{H}_k$, and $\langle \cdot, \cdot \rangle_{\mathcal{H}_k}$ and $\|\cdot\|_{\mathcal{H}_{k}}$ for its inner product and norm, respectively. Associated to each RKHS, there exists a symmetric and positive definite function $k:\mathbb{R}^d \times \mathbb{R}^d \rightarrow \mathbb{R}$ called a reproducing kernel \cite{Berlinet2004}. This function satisfies two properties: (i) for all $f\in \mathcal{H}_k$, $f(\mathbf{x})=\langle f, k(\mathbf{x},\cdot)\rangle_{\mathcal{H}_k}$ (called the reproducing property), and (ii) $k(\mathbf{x},\cdot) \in \mathcal{H}_k$ for all $\mathbf{x} \in \mathbb{R}^d$.

It is straightforward  to map probability distribution  $\mathbb{P}$ to $\mathcal{H}_k$ through what is called a \emph{kernel mean embedding} defined  as
\begin{align}
\label{eq:meanEmbedding}
    \mu_{\mathbb{P}}(\cdot) = \mathbb{E}_{X\sim \mathbb{P}}[k(X,\cdot)] = \int_{\mathbb{R}^d} k(\mathbf{x, \cdot}) \mathbb{P}(\text{d}\mathbf{x}),
\end{align}
under mild regularity conditions satisfied for all kernels in this paper, see \cite[Lemma 3]{Gretton2012JMLR}. Here,  $\mathbb{E}[\cdot]$ denotes the expectation with respect to the random variable and probability distribution given in subscript. Note that, $\mu_{\mathbb{P}} \in \mathcal{H}_k$. In the case where the probability distribution $\mathbb P$ has a probability density function $p$, the integral in \eqref{eq:meanEmbedding} can be written in the more wide-spread form $\int_{\mathbb{R}^d} k(\mathbf x,\cdot)p(\mathbf x) \text{d}\mathbf x $. Alternatively, when $\mathbb{P}$ is an empirical distribution corresponding to a data-set, then the kernel mean embedding is given by $\frac{1}{N_X} \sum_{i=1}^{N_X} k(\mathbf{x}_i,\mathbf{x})$. 

The MMD between probability distributions $\mathbb{P}$ and $\mathbb{Q}$ embedded in $\mathcal{H}_k$ is defined as the supremum taken over the mean of all functions in the unit ball in an RKHS, i.e. \cite{Muandet2017}
\begin{align}
\label{eq:MMDsupreme}
  \mathrm{MMD}_k[\mathbb{P},\mathbb{Q}] =   \sup_{\|f\|_{\mathcal{H}_{k}}\leq 1} \left| \mathbb{E}_{X \sim \mathbb{P}}[f(X)]- \mathbb{E}_{X \sim \mathbb{Q}}[f(X)] \right|.
\end{align}
As the name suggests, the MMD is the maximum distance between means of (unit norm) functions computed with respect to the distributions  $\mathbb{P}$ and $\mathbb{Q}$.
As shown in \cite{Gretton2012JMLR}, the MMD in \eqref{eq:MMDsupreme} can equivalently be expressed as  
\begin{align}\label{eq:MMD}
    \mathrm{MMD}_k[\mathbb{P}, \mathbb{Q}] &= \left\Vert \mathbb{E}_{X\sim \mathbb{P}}[k(X,\cdot)] - \mathbb{E}_{Y\sim \mathbb{Q}}[k(Y,\cdot)] \right\Vert_{\mathcal{H}_k} \\
    &= \left\Vert \mu_\mathbb{P} - \mu_\mathbb{Q} \right\Vert_{\mathcal{H}_k} \nonumber
\end{align}
This gives  an alternative interpretation of the MMD as the distance between mean embeddings in $\mathcal{H}_k$ as  Fig.~\ref{fig:embedding} illustrates. 

A third expression for the MMD appears upon expanding the squared norm in \eqref{eq:MMD} and using the reproducing property of $k$ which yields an expression in terms of $k$ as
\begin{multline}\label{eq:mmdSquared}
    \mathrm{MMD}_k^2[\mathbb{P}, \mathbb{Q}] = \mathbb{E}_{X,Y \sim \mathbb{P}}[k(X, Y)]\\
    - 2\mathbb{E}_{X \sim \mathbb{P}, Y \sim \mathbb{Q}}[k(X, Y)] + \mathbb{E}_{X,Y \sim \mathbb{Q}}[k(X, Y)].
\end{multline}
The latter expression is computationally more appealing than the two former as it only calls for computation of expectations of the kernel. Thus, computation of the supremum in \eqref{eq:MMDsupreme} is not required to compute the MMD. As discussed later in Section~\ref{sec:MMDfromdata}, the expression \eqref{eq:mmdSquared} forms the basis for estimation of the MMD from data.

\subsection{Selecting a Kernel}
The choice of kernel defines the RKHS and thus the properties of its distance, the MMD. In addition to being reproducing, it is a great advantage if the kernel is \textit{characteristic} \cite{Sriperumbudur2009,Simon-Gabriel2016}.  This implies that the kernel mean embedding is an injective mapping, meaning that each distribution is mapped to a unique function. Thus, in the case of characteristic kernels, the kernel mean embedding captures all the information about the distribution. As a result, $\mathrm{MMD}_k[ \mathbb{P}, \mathbb{Q}]  = \Vert \mu_\mathbb{P} - \mu_\mathbb{Q} \Vert_{\mathcal{H}_k} = 0$ if and only if $\mathbb{P} = \mathbb{Q}$. In this case, the MMD is capable of comparing infinitely many moments of two probability distributions without ever having to compute these moments explicitly.  Consequently, the MMD is able to distinguish probability distributions even when these coincide in finite number of moments. This gives a great advantage over methods based on comparison of finitely many moments which are potentially blind to differences between distributions.

A very popular characteristic reproducing kernel is the squared-exponential (or Gaussian) kernel, defined as
\begin{equation}\label{eq:gaussianKernel}
    k_{\text{SE}}(\mathbf{x}, \mathbf{x}') = \exp\left(- \frac{\Vert \mathbf{x} - \mathbf{x}' \Vert_2^2}{l^2} \right),  
\end{equation}
for $\mathbf{x}, \mathbf{x}' \in \mathbb{R}^d$. Here, $\|\cdot\|_{2}$ is the Euclidean norm and $l>0$ is a parameter called the lengthscale of the kernel. The norm inside the exponent can be chosen based on the specific data and application. 
For additional examples of characteristic kernels, see \cite{Sriperumbudur2009,Simon-Gabriel2016}.

We now give a simple example comparing Gaussian distributions, in which case the MMD can be derived analytically. 
\subsubsection*{Example} 
\begin{figure}
	 \centering
		 \includegraphics[trim={125 325 130 338}, clip, width = \columnwidth]{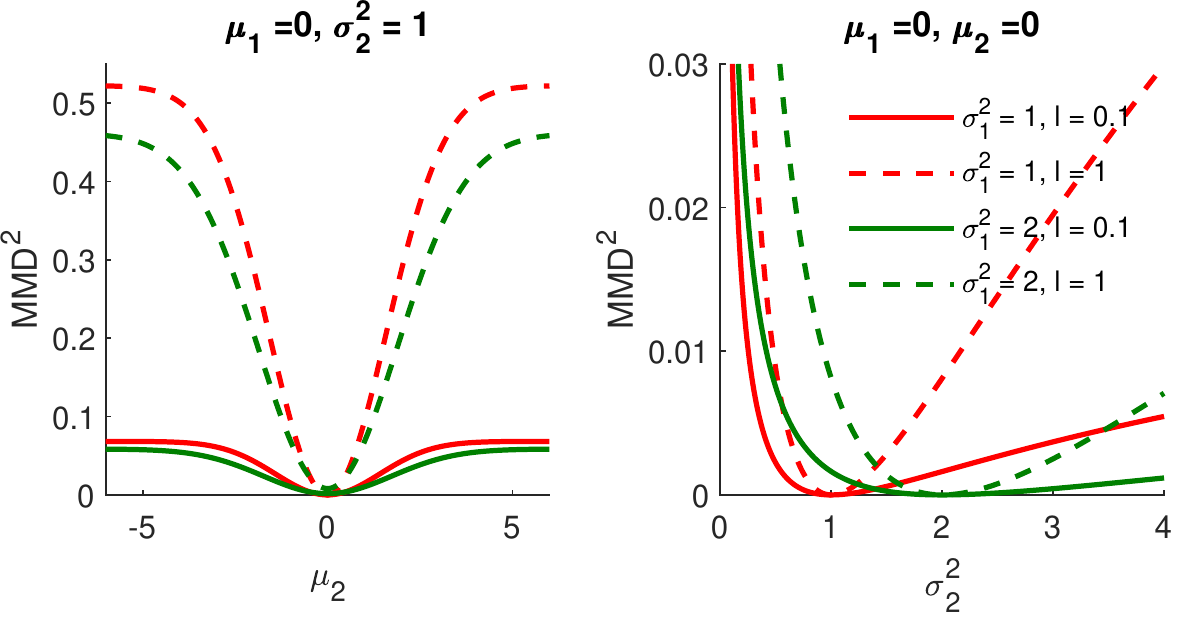}
	 \caption{MMD between a $\mathcal{N}(\mu_1,\sigma_1^2)$ distribution with $\mu_1 = 0$ and $\sigma_1 = 1$ and a $\mathcal{N}(\mu_2,\sigma_2^2)$ varying parameters $\mu_2$ and $\sigma_2$. Plots are shown for two different values of the lengthscale $l$.}
	 \label{fig:toy1} 
 \end{figure}

Let $\mathbb{P} = \mathcal{N}(\mu_1, \sigma_1^2)$ and $\mathbb{Q} = \mathcal{N}(\mu_2, \sigma_2^2)$ be two Gaussian distributions on $\mathbb R$. For the squared-exponential kernel in \eqref{eq:gaussianKernel}, the MMD takes the form (see \cite[Appendix~C]{Briol2015}):
\begin{multline}\label{eq:theoreticalMMD}
    \mathrm{MMD}_{k_{\text{SE}}}^2[ \mathbb{P}, \mathbb{Q}] = \frac{l}{l + 2\sqrt{2}\sigma_1} +\frac{l}{l + 2\sqrt{2}\sigma_2} \\
    - \frac{2l}{l + \sqrt{2}\sigma_1 +\sqrt{2}\sigma_2} \exp{\left(- \frac{(\mu_1 - \mu_2)^2}{l^2 + 2\sigma_1^2 + 2\sigma_2^2} \right)}.
\end{multline}
It is apparent from \eqref{eq:theoreticalMMD} that the $\text{MMD}$ is zero if and only if $\mu_1=\mu_2$ and $\sigma_1 = \sigma_2$ (as guaranteed by using a characteristic kernel). Fig.~\ref{fig:toy1} illustrates how the $\text{MMD}$ increases as the parameters of these distributions increasingly differ. Varying the lengthscale, $l$, of the kernel scales the overall MMD curve, but does not affect the point at which the MMD is minimised. The overall behaviour of the curves do not vary significantly on changing the lengthscale by an order of the magnitude.

\subsection{Maximum Mean Discrepancy Between Data-sets}\label{sec:MMDfromdata}

Unlike in the previous example, it is, in most realistic cases, not feasible to analytically calculate \eqref{eq:mmdSquared}. Moreover, numerical integration is problematic, as the dimension of $X$ and $Y$ may be large and $\mathbb{P}$ or $\mathbb{Q}$ unavailable. Fortunately, it is straightforward to estimate the MMD if it is an empirical distribution, such as in the case of data-sets. 

Imagine that we do not have access to $\mathbb{P}$ and $\mathbb{Q}$, but that we instead have two data-sets consisting of realisations from these distributions. More precisely, suppose we have access to  $\mathbf{X} = \{\mathbf{x}_1, \dots, \mathbf{x}_{N_X}\} \overset{iid}{\sim} \mathbb{P}$ and $\mathbf{Y} = \{\mathbf{y}_1, \dots, \mathbf{y}_{N_Y}\} \overset{iid}{\sim} \mathbb{Q}$. Then, an unbiased empirical estimate of $\mathrm{MMD}^2_k[\mathbb{P}, \mathbb{Q}]$ can be obtained as \cite{Gretton2012JMLR}
\begin{multline}\label{eq:empiricalMMD}
             \widehat{\mathrm{MMD}}_k^2[\mathbf{X}, \mathbf{Y}] = \frac{\sum_{i\neq i'} k(\mathbf{x}_i, \mathbf{x}_{i'})}{N_X(N_X-1)} \\
            - \frac{2\sum_{j=1}^{N_Y} \sum_{i=1}^{N_X} k(\mathbf{x}_i, \mathbf{y}_j)}{N_Y N_X}    + \frac{\sum_{j\neq j'} k(\mathbf{y}_j, \mathbf{y}_{j'})}{N_Y(N_Y-1)}.
\end{multline}
Note that $N_X$ and $N_Y$ are permitted to differ, i.e. the two data-sets are not limited to be of the same size. To use this estimator with the kernel in \eqref{eq:gaussianKernel}, the lengthscale should be specified. Following \cite{Gretton2012JMLR}, the lengthscale can be set based on the data-set $\mathbf X$  using the median heuristic
\begin{equation}\label{eq:lengthscale}
    l~=~\sqrt{\mathrm{med}/ 2},
\end{equation} 
where $\mathrm{med}$ denotes the median of the set of squared two-norm distances $ \Vert \mathbf{x}_i - \mathbf{x}_j \Vert_2^2$ for all pairs of distinct data points in $\mathbf X$. This setting of $l$ scales the kernel with the spread of the data, and is robust to outliers.

Concentration bounds for MMD, such as \cite[Lemma 1]{Briol2019MMD} or  \cite[ Theorem 3.4]{Muandet2017}, imply that with high probability, 
\begin{align}
\left|\widehat{\mathrm{MMD}}_k^2[\mathbf{X}, \mathbf{Y}] - \mathrm{MMD}_k^2[\mathbb{P}, \mathbb{Q}]\right| \leq C \left(\frac{1}{N_X}+ \frac{1}{N_Y}\right),
\end{align}
for some $C>0$. This tells us that the accuracy of the estimate converges linearly in both $N_X$ and $N_Y$. The computational cost of computing this estimate is $\mathcal{O}(N_X^2+N_Y^2)$ due to the need to compute double sums in both $N_X$ and $N_Y$. In order to best balance computational cost and accuracy, $N_X$ and $N_Y$ should be chosen to be commensurate. These two results on accuracy and computational cost can be used to determine how to make default choices for the parameters of our ABC algorithm.
\begin{figure*}
	 \centering
	 \scalebox{0.75}{
	% Define block styles
\tikzstyle{decision} = [diamond, draw, fill=blue!90, 
    text width=4.5em, text badly centered, node distance=3cm, inner sep=0pt]
    % cyan!30, 
\tikzstyle{block} = [rectangle, draw, fill=white!100, 
    text width=5em, text centered, rounded corners, minimum height=4em]
\tikzstyle{line} = [draw, -latex', line width = 1pt]
\tikzstyle{cloud} = [draw, ellipse,fill=red!20, node distance=3cm,
    minimum height=2em]
\tikzstyle{pointt} = [circle, draw, fill=black, minimum size=#1,
              inner sep=2pt, outer sep=2pt]
 \tikzstyle{output} = [coordinate]

\begin{tikzpicture}[node distance = 2.5cm, auto]

\large
% Place nodes
\draw[dashed,draw=black] (7.75,-1.25) rectangle ++(9,3);
\node [above] at (16.8,1.8) {Alg.~\ref{alg:regressionABC}};

\node [block] (prior) {Sample from Prior};
\node [block, right of=prior, node distance=4.5cm] (model) {Simulate from model };
\node [block, right of=model, node distance=5cm] (rej) {Rejection based on MMD};
\node [block, right of= rej, node distance=5.5cm] (regression) {Regression adjustment};
\node [above of=rej, node distance = 2.5cm] (obsData) {};
\node [right of=obsData, node distance=2.75cm] (data) {Data $ \mathbf{Z} $};
 \node [ below of = data, node distance = 1cm] (temp) {};
% \node [block, right of=rej, node distance=4.5cm] (summarize) {Compute means and covariances};
\node [output, right of = regression, node distance=3.5cm] (output) {};
\node [block, below of=regression, text width=7em] (pmc) {Sample from importance distribution};

\path [line] (prior) -- node {$\Theta$} (model);
\path [line] (model) -- node {$(\mathcal{X}, \Theta)$}(rej);
\path [line] (rej) -- node  {$(\mathcal{X}^*, \Theta^*)$} (regression);
 \path [line] (regression) -- node [name=y] {$\tilde{\Theta}$} (output);
 \path [line] (y) |-   (pmc);
\path [line] (pmc) -| node [near start] {$\Theta$} (model);
\path [line] (data) -| (regression);  
\path [line] (data) -| (rej);  

\end{tikzpicture}}
	 \caption{Diagram depicting steps in the proposed kernel-based ABC algorithm with regression adjustment described in Alg.~\ref{alg:pmc-abc}. The block ``Rejection based on MMD'' corresponds to Sec.~\ref{sec:rejection}, ``Regression adjustment'' corresponds to Sec.~\ref{sec:adjustment}, and ``Sample from importance distribution" corresponds to Sec.~\ref{sec_newPopulation}. Here the term ``Data" can be either obtained from physical measurements or as in Sec.~\ref{sec:5} by simulation.}
	 \label{fig:PMC-ABC}
 \end{figure*}
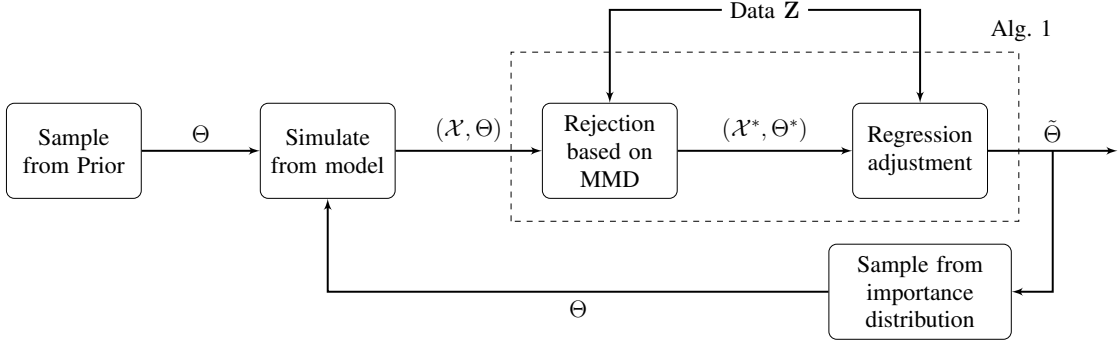

\subsection{Kernels for Radio Channel Measurements}

In order to use the MMD for calibrating stochastic radio channel models, we need a kernel defined on the space of transfer function measurements: $k_{\mathcal{Y}}:\mathcal{Y}\times \mathcal{Y} \rightarrow \mathbb{R}$. Given such a kernel, we could then estimate the MMD between a measured data-set $\mathbf{Y}$ and a data-set $\mathbf{Y}_{\mathrm{sim}}$ simulated from the model. 

A significant challenge with this approach is that, in the context of stochastic radio channel models, $\mathcal{Y}$ is usually a high-dimensional space. This is especially the case for large bandwidth measurements where $N_s$ can be in the order of thousands. Such high-dimensional problems are challenging for kernel methods based on default kernels such as the squared-exponential kernel \cite{Reddi2015}. These kernels indeed suffer from the curse-of-dimensionality, a phenomenon implying that the distance between points increases exponentially with the dimension of the space. 

To tackle this issue, there exist kernels specialised to certain time-series or functional data models in the literature \cite{Ruping2001, Cuturi2007,Cuturi2011A, Chevyrev2018,Kiraly2019, Wynne2020}. These use specific properties of the type of data in order to avoid the curse-of-dimensionality. In this paper, we contribute to this literature and construct a kernel specifically tailored to transfer function measurements.
We base the kernel on the temporal moments of $y(t)$, defined as
\begin{equation}\label{eq:temporalMoments}
    m^{(i)} = \int_0^{t_{\mathrm{max}}} t^i \vert y(t)\vert^2 \text{dt} , \quad i = 0,1,2,\dots, I.
\end{equation}
The integral in \eqref{eq:temporalMoments} is easy to compute numerically. The temporal moments can be seen as an expansion of $|y(t)|^2$ into the basis of monomials. Since the monomials form a complete basis for finite energy time-limited signals \cite{Franks1969}, no information is lost compared to $|y(t)|^2$ if  $I\rightarrow \infty$. Referring to \cite{AyushEuCAP, Bharti2021}, the first few moments are well modeled by a log-normal distribution. Thus, taking the entry-wise logarithm $ z^{(i)} = \ln m^{(0)}$ brings the moments to the same scale and gives an approximately Gaussian vector $\mathbf{z} = [z^{(0)}, \dots, z^{(I-1)}]$. Multiple channel realizations yield $\mathbf{Z} = \left(\mathbf{z}_1, \mathbf{z}_2, \dots, \mathbf{z}_{\nreal} \right)$.

Define the mapping $A_I: \mathcal{Y} \rightarrow \mathbb{R}^I$ from $\mathcal{Y}$ to the $I$-dimensional space of log temporal moments. We propose to construct a kernel $k_{\mathcal{Y}}$ for transfer function data as 
\begin{equation}
    k_{\mathcal{Y}}\left( \mathbf{y}, \mathbf{y}' \right) := k_{\text{SE}}\left( A_I(\mathbf{y}), A_I(\mathbf{y}') \right),
    \quad \text{for all } \mathbf{y},\mathbf{y}' \in \mathcal{Y},
\end{equation}
where $k_{\text{SE}}$ denotes the squared-exponential kernel in dimension $I$. We note that this is the composition of a reproducing kernel and a map, and thus according to \cite[Lemma 4.3]{Steinwart2008} is a reproducing kernel on $\mathcal{Y}$. We also note that the MMD with kernel $k_{\mathcal{Y}}$ computed on the original data can be obtained through the MMD with kernel $k_{\text{SE}}$ on the log temporal moments. Similarly, the empirical estimators of these quantities are also identical, i.e.
\begin{align}\label{eq:MMDky_is_MMDkSE}
    \widehat{\mathrm{MMD}}^2_{k_\mathcal{Y}}[\mathbf{Y},\mathbf{Y}_{\text{sim}}] = \widehat{\mathrm{MMD}}^2_{k_{\text{SE}}}[\mathbf{Z},\mathbf{X}],
\end{align}
where $\mathbf{X}$ is the simulated log temporal moments data-set.

In practice, we will have to limit ourselves to a finite $I$ for computational reasons. This, however, is not a problem since we can expect the signal energy to be concentrated on the lowest moments. In fact, taking $I$ to be small also allows us to by-pass issues with the curse-of-dimensionality.

From a theoretical viewpoint, since the squared-exponential kernel is characteristic, we should be able to recover any distribution on the space of log temporal moments. However, since the mapping $A_I$ leads to loss of information when $I$ is finite, $k_{\mathcal{Y}}$ will not be characteristic on $\mathcal{Y}$, and we may not be able to uniquely identify the distribution on $|y(t)|^2$. This however is not an issue for the considered channel models, as will be shown in Section \ref{sec:5}.

\section{Proposed Kernel-based Approximate Bayesian Computation Method}\label{sec:4}

ABC methods rely on simulation from the model to approximate the posterior, and can be used to estimate $\btheta$ such that the model fits to the observed data $\mathbf{Y}$. Let $\rho(\cdot, \cdot)$ be some notion of distance between data-sets. The basic form of ABC, called rejection ABC, proceeds by sampling $M$ parameter values from $p(\btheta)$ and generating the corresponding simulated data $\ysim$ from the model. The values of $\btheta$ for which $\rho(\mathbf{Y}, \ysim)$ is less than some pre-defined threshold $\epsilon$, form a sample from the approximate posterior distribution, $\tilde{p}(\btheta | \mathbf{Y}) = p(\btheta | \rho(\mathbf{Y}, \ysim) < \epsilon)$. The tolerance threshold impacts the degree of approximation in ABC methods. Setting $\epsilon = 0$ would lead to exact Bayesian inference, however, achieving equality for continuous-valued data is not possible. Hence, $\epsilon$ should be small but non-zero in order to be computationally feasible.

We now propose an ABC method based on the MMD as the distance metric to calibrate stochastic radio channel models. We employ the Population Monte Carlo (PMC) ABC method \cite{Beaumont2009} to iteratively refine our approximation of the ABC posterior. At the end of each iteration, we perform  local-linear regression adjustment \cite{Beaumont2002} to further improve the posterior approximation. The complete algorithm is depicted in Fig.~\ref{fig:PMC-ABC} and outlined in Alg.~\ref{alg:pmc-abc}. Individual steps of this PMC-ABC algorithm will be highlighted in Sec.~\ref{sec:rejection} to \ref{sec_newPopulation}. In Sec.~\ref{sec:mismatch}, we describe how to detect and account for model misspecification in the algorithm.

\subsection{Rejection based on MMD}\label{sec:rejection}

The proposed ABC method uses the MMD between data-sets as a rejection criteria. Instead of setting the threshold $\epsilon$ in terms of the distance, we specify the proportion of accepted samples, i.e. $\epsilon = M_\epsilon/M$ where $M_\epsilon$ is the number of parameter samples accepted out of $M$. This is particularly convenient as it avoids the need to manually find a threshold, which may lead to unknown run-time of the algorithm. 

The method computes $\widehat{\mathrm{MMD}}_{k_\text{SE}}^2[\mathbf{X}, \mathbf{Z}]$, where $\mathbf{X} = \left(\mathbf{x}_1 ,\dots, \mathbf{x}_{\nsim}\right)$ is the simulated log temporal moments data-set, as this is identical to estimating the MMD between $\mathbf{Y}$ and $\ysim$ (see Eq. \ref{eq:MMDky_is_MMDkSE}). First, $M$ independent parameter samples $\Theta = (\btheta_1, \dots, \btheta_M)$ are drawn from the prior $p(\btheta)$. For each $\btheta_i$,  the log temporal moments data-set, $\mathbf X_i  \sim \mathbb{P}_{\btheta_i}$, is simulated. The simulated data-sets are gathered in $\mathcal{X} = \left(\mathbf{X}_1, \dots, \mathbf{X}_M \right)$. The $\widehat{\mathrm{MMD}}_{k_\text{SE}}^2[\mathbf{X}_i, \mathbf{Z}]$ is computed for each $i$ using \eqref{eq:empiricalMMD}, setting the lengthscale of $k_\text{SE}$ as per $\eqref{eq:lengthscale}$. The parameter samples resulting in the $M_\epsilon$ smallest MMD values are then accepted.

In principle, the MMD could be computed between the samples of the temporal moments instead of their logarithm. However, the magnitudes of the different temporal moments may vary strongly and using a single lengthscale may lead to poor performance. Using a log transformation helps mitigate this issue. Alternatively, the lengthscale should be defined for each dimension of $\btheta$.

\subsection{Regression Adjustment}\label{sec:adjustment}

As proposed in \cite{Beaumont2002}, it is possible improve the posterior approximation by adjusting the accepted samples using a model of the relationship between a low-dimensional vector of statistics and the parameter vector. Let $\mathbf{s}$ be a vector of summary statistics of $\mathbf{X}$ such that $\mathbf{s} = S(\mathbf{X})$ for a function $S(\cdot)$. Similarly, the observed summary statistics are denoted $\sobs = S(\mathbf{Z})$. We begin by fitting a function, $g$, between the accepted parameters $\Theta^* = (\btheta_1^*, \dots, \btheta_{M_\epsilon}^*)$ and the corresponding statistics $\mathcal{S}^* = (\mathbf{s}_1, \dots, \mathbf{s}_{M_\epsilon})$ as \cite[Ch. 3]{Sisson2018}

\begin{equation}
    \btheta_i = g(\mathbf{s}_i) + \varepsilon, \quad i = 1, \dots, M_\epsilon,
\end{equation}
where $g(\mathbf{s})$ is the conditional expectation of $\btheta$ given $\mathbf{s}$, and $\varepsilon$ is the residual. Here, $\btheta$ should belong to a subset of $\mathbb{R}^p$. Considering that the log of the temporal moments are well modeled by a Gaussian distribution, we take $\sobs$ to be the vector consisting of the sample means and sample covariances of the elements of $\mathbf{z}$, similar to \cite{Bharti2020}. In total, $\sobs$ consists of $(I^2 + 3I)/2$ elements for $I$ temporal moments. The statistics $\mathbf{s}$ is computed in the same manner for $\mathbf{X}$. Note that $\mathbf{s}$ and $\sobs$ are normalized by an estimate of their median absolute deviation to account for the difference in magnitude of the statistics. In case the prior distributions are bounded, a logit transformation is applied to the parameters before the adjustment.

For simplicity reasons, we assume $g$ to be linear as in \cite{Beaumont2002} and adjust the accepted parameters as
\begin{equation}\label{eq:adjustedSamples}
    \tilde{\boldsymbol{\theta}}_i = \boldsymbol{\theta}_i^* -  \left(\mathbf{s}_i - \mathbf{s}_{\text{obs}} \right)^\top \hat{\boldsymbol{\beta}},\quad i = 1,\dots, M_\epsilon,
\end{equation}
where $\hat{\boldsymbol{\beta}}$ is the solution to the weighted least-squares problem
\begin{equation}\label{eq:objFunction}
    \underset{\boldsymbol{\alpha},\boldsymbol{\beta}}{\arg \min}\sum_{i=1}^{M_\epsilon} \left[\boldsymbol{\theta}_i^* - \boldsymbol{\alpha} - \left(\mathbf{s}_i - \mathbf{s}_{\text{obs}}\right)^\top \boldsymbol{\beta} \right]^2 \mathcal W_{\left(\widehat{\mathrm{MMD}}_{k_\text{SE}}^2[\mathbf{X}_i, \mathbf{Z}]\right)}.
\end{equation}
The weighting function $\mathcal W$ applies weights to each $\btheta_i$ based on the estimated MMD value. This guarantees that parameters which yield simulated log moments ``closer to" $\mathbf{Z}$ are weighted more heavily. We take $\mathcal W$ to be the Epanechnikov function, $\mathcal W_{(\delta)} = 1 - (\delta / \delta_{\mathrm{max}})^2$ for $|\delta| \leq \delta_{\mathrm{max}}$ and zero otherwise, as proposed in \cite{Beaumont2002}. Here, $\delta_{\mathrm{max}}$ is the maximum estimated MMD associated to the accepted parameters. Note that choosing a constant regression function, i.e. $\boldsymbol \beta = 0$, and assigning equal weights to all $\btheta_i$'s results in the basic rejection ABC algorithm. The regression adjustment therefore gives the adjusted parameter values $ \tilde{\Theta} = (\tilde{\btheta}_1, \dots, \tilde{\btheta}_{M_\epsilon})$. 

\begin{algorithm}[t]
	\caption{ABC with MMD and Regression Adjustment}
	\label{alg:regressionABC}
	\small
	\textbf{Input}: Parameter values $\Theta$, corresponding simulated data $\mathcal{X}$, observed $\mathbf{Z}$ \& number of accepted samples $M_\epsilon$.
	\begin{algorithmic}
		\STATE Compute $\widehat{\mathrm{MMD}}_{k_\text{SE}}^2(\mathbf{X}_i, \mathbf{Z})$ for all data-sets $\mathbf{X}_i \in \mathcal{X}$ using \ref{eq:empiricalMMD}.
		\STATE Accept the $M_\epsilon$ parameters with the smallest MMD distance and denote these $\Theta^* = (\btheta^*_1, \dots, \btheta^*_{M_\epsilon})$.
		\STATE Compute $\mathcal{S}^*$ and $\sobs = S(Z)$, then solve the optimisation problem in \eqref{eq:objFunction} with $\Theta^*$, $\mathcal{S}^*$, and $\sobs$ to get $\hat{\boldsymbol{\beta}}$. 
		\STATE Adjust $\Theta^*$ using \eqref{eq:adjustedSamples} to obtain $\tilde{\Theta}$.
	\end{algorithmic}
	\textbf{Output}: Adjusted samples $\tilde{\Theta}$ from the Rejection-ABC posterior.
	
\end{algorithm}
\begin{figure}
\centering
  \includegraphics[trim={75 180 55 20}, clip, width = \columnwidth]{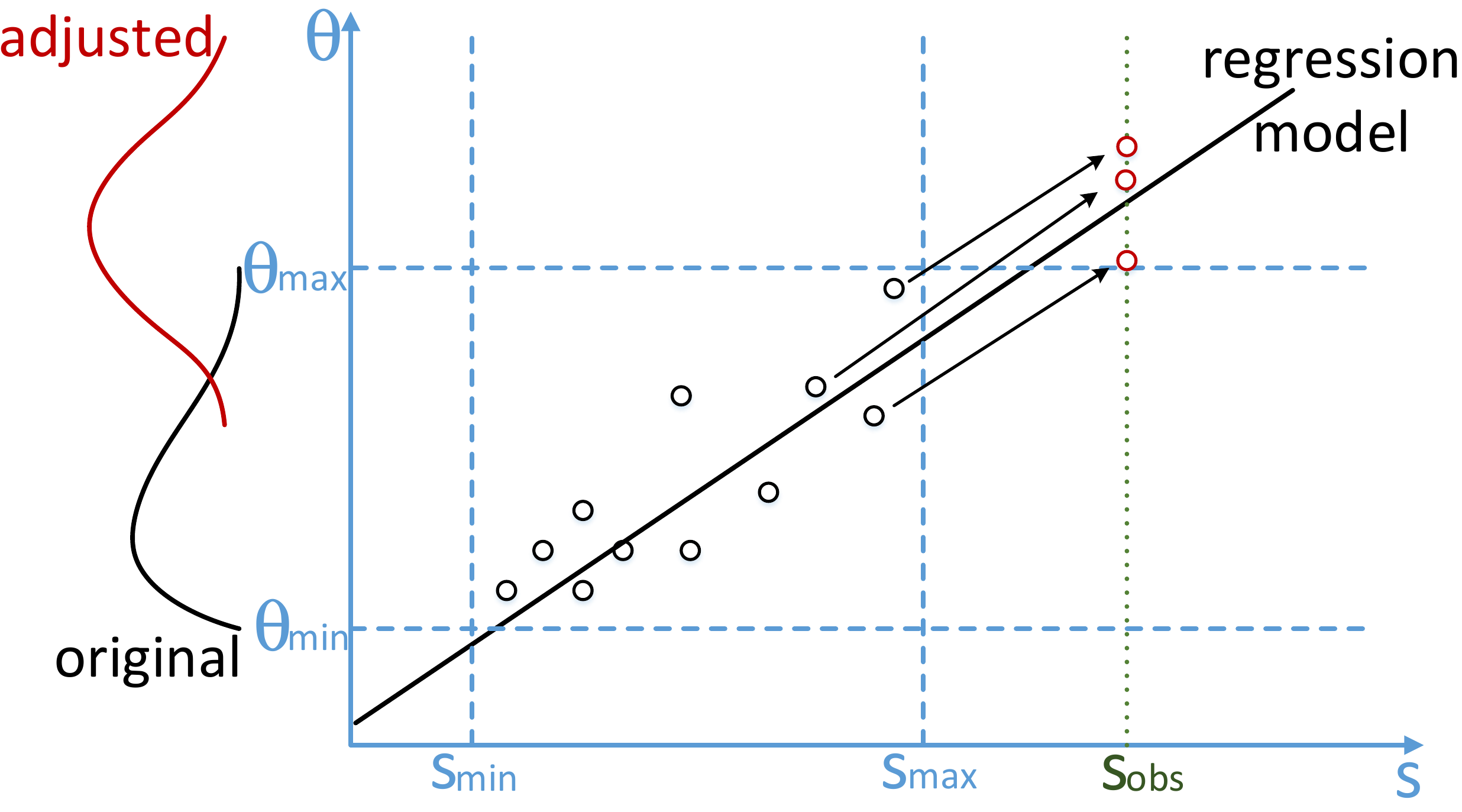}
  \caption{Local linear regression adjustment of parameter $\theta$ inspired from \cite{Lintusaari2016}. First, the regression model is fitted based on accepted parameter and statistic values. Then, the parameters are adjusted based on the fitted model, which can move them outside the prior range if $(s_i - s_{\mathrm{obs}})$ is large.}
      \label{fig:mismatch}
\end{figure}%

\subsection{Importance Sampling using PMC} \label{sec_newPopulation}

As a means to explore the posterior distribution over the parameter space efficiently, we employ a sequential Monte Carlo technique called PMC \cite{Beaumont2009, AyushABC, Bharti2020}. In PMC, the current parameter values $\tilde{\Theta}$ are used to generate a new set of parameters for the next iteration of the algorithm through importance sampling. This is a two-step procedure: (1) sample from the current parameters based on their importance weights, and (2) perturb the sampled parameter values using a proposal density.

The set of parameters in the initial iteration,  $\tilde{\Theta}^{(1)} = (\tilde{\btheta}_1^{(1)}, \dots, \tilde{\btheta}_{M_\epsilon}^{(1)})$, are assigned equal weights. The next set of parameters is obtained by drawing $M$ values from $\tilde{\Theta}^{(1)}$ and perturbing these according to a probability distribution, called proposal. For simplicity, we perturb independently in each dimension using a Gaussian distribution, and reject values outside the prior range. Thus, the proposal reads
\begin{equation}
 \varphi
 (\btheta ; \tilde{\btheta}, \boldsymbol{\Sigma}) =
  \mathbb{1}(\btheta \in \mathcal{R})  e^{- \frac{1}{2}
  (\btheta-\tilde\btheta)^\top\boldsymbol\Sigma^{-1}   (\btheta-\tilde\btheta) }
\end{equation}
where $\mathbb{1}$ is an indicator function, $\mathcal R \subset \mathbb{R}^p$ is the prior range, and $\boldsymbol\Sigma$ is a diagonal matrix with variances $\sigma^2_j>0$ corresponding to parameter $\theta_j$ along the diagonal. We set the diagonal elements of $\boldsymbol \Sigma$ to twice the empirical variance of the adjusted parameter samples. This is denoted as $\boldsymbol \Sigma = 2 \widehat{\mathrm{Var}}( \tilde \Theta)$.

The set of $M$ parameter values at iteration $t$, ${\Theta}^{(t)}$, is then used to simulate $\mathcal{X}^{(t)}$ from the model for MMD computation and regression adjustment (i.e. Alg.~\ref{alg:regressionABC}). In subsequent iterations, weights are assigned as 
\begin{equation}\label{eq:weights} 
     w_j^{(t)} \propto p\left(\btheta_j^{(t)}\right) / \sum_{i=1}^{M_\epsilon} w_i^{(t-1)} \varphi \left( \btheta_j^{(t)}; \tilde{\btheta}_i^{(t-1)}, \boldsymbol{\Sigma}^{(t-1)}\right) ,
\end{equation}
$j=1,\ldots, M_\epsilon$. The adjusted parameter values after iteration $T$ are taken as samples from the approximate posterior distribution. Point estimates of $\btheta$, such as the approximate posterior mean, 
\begin{equation}
    \hat{\btheta}^{(T)} = \frac{1}{M_\epsilon} \sum_{i=1}^{M_\epsilon} \tilde{\btheta}_i^{(T)},
\end{equation}
are straightforward to compute from the samples.
\begin{algorithm}[t]
	\caption{PMC-ABC with MMD}
	\label{alg:pmc-abc}
	\small
	\textbf{Input}: Prior $p(\btheta)$, model $\mathbb{P}_{\btheta}$, observed data $\mathbf{Z}$, $M_\epsilon$, $M$ and $T$.
	\begin{algorithmic}
		\STATE Initialize $t=1$, draw $\Theta^{(1)} \overset{\text{iid}}{\sim} p(\btheta)$ and simulate $\mathcal{X}^{(1)}$ using the parameters in $\Theta^{(1)}$.
		\STATE Apply \textbf{Algorithm} \ref{alg:regressionABC} on $\{\mathcal{X}^{(1)}, \Theta^{(1)}\}$ to obtain $\tilde{\Theta}^{(1)}$.
		\STATE Set $w_j^{(1)} = 1$ for $j = 1, \dots, M_\epsilon$, and set $\boldsymbol{\Sigma}^{(1)} = 2 \widehat{\mathrm{Var}}\big(\tilde{\Theta}^{(1)} \big)$.
	    \FOR{$t = 2,\ldots,T$}
		  \STATE Compute $q_j = w_j^{(t-1)} /\sum_{i=1}^{M_\epsilon} w_i^{(t-1)}$ for $j=1,\ldots,M_{\epsilon}$.
		   \FOR{$i = 1, \dots , M$}
		  \STATE Sample $\btheta_i^*$ from $\tilde{\Theta}^{(t-1)}$ s.t. $\tilde{\btheta}_j^{(t-1)}$ is selected with prob $q_j$.
			\STATE Generate $\btheta_i^{(t)} \sim \varphi\big( \cdot; \btheta_i^*,\boldsymbol{ \Sigma}^{(t-1)} \big)$. 
			\STATE Simulate $\mathbf{X}_i^{(t)}$ from the model with parameter $\btheta^{(t)}_i$.
		    \ENDFOR
		\STATE Apply \textbf{Algorithm} \ref{alg:regressionABC} on $\{\mathcal{X}^{(t)}, \Theta^{(t)}\}$ to obtain $\tilde{\Theta}^{(t)}$.
		\STATE Set $w_j^{(t)}$ using \eqref{eq:weights} for $j=1,\ldots, M_\epsilon$.
		\STATE Set $\boldsymbol{\Sigma}^{(t)} = 2 \widehat{\mathrm{Var}}\big(\tilde{\Theta}^{(t)}\big)$.
		\ENDFOR
	\end{algorithmic}
	\textbf{Output}: Samples $\big(\tilde{\btheta}_1^{(T)}, \dots, \tilde{\btheta}_{M_\epsilon}^{(T)}\big)$ from the PMC-ABC posterior. 
\end{algorithm}

\subsection{Handling Model Misspecification} \label{sec:mismatch}

We have now completed the description of Alg.~\ref{alg:pmc-abc}. However, the framework of ABC relies on the implicit assumption that there exist parameter values in the prior support that yield simulated data ``close" to the measured data. This assumption may not always hold if the model parameters cannot be set in any way to reproduce the data well. In this case, we say that the model is misspecified for the data. Misspecification can be detected and accounted for in the algorithm as explained in this subsection.

Consider a univariate parameter $\theta$ in the range $[\theta_{\mathrm{min}}, \theta_{\mathrm{max}}]$ resulting in a univariate statistic $s$ in $[s_{\mathrm{min}}, s_{\mathrm{max}}]$ simulated from the model. If the observed statistic $s_{\mathrm{obs}} \notin [s_{\mathrm{min}}, s_{\mathrm{max}}]$, then the model is likely to be misspecified. This is a challenge since under model misspecification, the local-linear regression adjustment has been shown to concentrate posterior mass on a completely different value than the rejection ABC \cite{Frazier2020}. In fact for parameters with bounded support, the regression adjustment moves the parameter samples outside the prior range as illustrated in Fig.~\ref{fig:mismatch}. Hence, if $s_{\mathrm{obs}}$ lies outside the range of statistics that the model can simulate, then there is no guarantee that the adjusted samples of $\theta$ will lie inside the prior range.

We check for model misspecification by observing whether each element of $\sobs$ lies within the range of corresponding statistics simulated from the model using $\Theta^{(1)}$. If any element of $\sobs$ lies outside the range of values simulated from the model, then the model is deemed misspecified. In such a case, we replace $\sobs$ by an alternative term, $\breve{\mathbf{s}}_{\mathrm{obs}}$, computed from the model instead of the data using the parameter  
\begin{equation}\label{eq:mismatchTheta}
    \breve{\btheta} = \underset{\btheta}{\arg \max} \quad f(\btheta;\Theta^*),
\end{equation}
where $f(\btheta;\Theta^*)$ is the kernel density estimate computed from the samples $\Theta^*$, and $\breve{\btheta}$ is the parameter corresponding to the mode of $f(\btheta;\Theta^*)$. Another choice for $\breve{\btheta}$ could be the posterior mean of rejection ABC \cite{Frazier2020}. However, we found the mean estimate to be unstable, especially in the initial iterations of the algorithm. Hence, in case of model misspecification, we set $\sobs = \breve{\mathbf{s}}_{\mathrm{obs}}$ in each iteration of the PMC-ABC algorithm, thus ensuring that the adjustment does not lead to parameter samples outside the prior range.

\section{Simulation Experiments}\label{sec:5}

We test the performance of the proposed calibration method on two different channel models, namely the Saleh-Valenzuela (S-V) and the propagation graph (PG) model. We chose models which differ significantly in their mathematical structure to highlight the generality of our approach. We first study in depth the advantages and drawbacks of our algorithm on simulated data. Then, in Section \ref{sec:6}, we calibrate these models to data from an indoor measurement campaign \cite{Gustafson2016}.

For ease of comparison, we use the same measurement settings as in \cite{Gustafson2016} for both simulations and measurements, i.e. $B =4$~GHz, $N_s = 801$, and $\tmax =200$~ns. We map the channel measurements to the first $I=4$ temporal moments. In each iteration of the ABC algorithm, $M = 2000$ parameter samples are generated, out of which $M_\epsilon = 100$ are accepted to estimate the posterior distribution.

\subsection{Application to the Saleh-Valenzuela model} 

\begin{figure}
\centering
  \includegraphics[trim={0 8 0 50}, clip, width = \columnwidth]{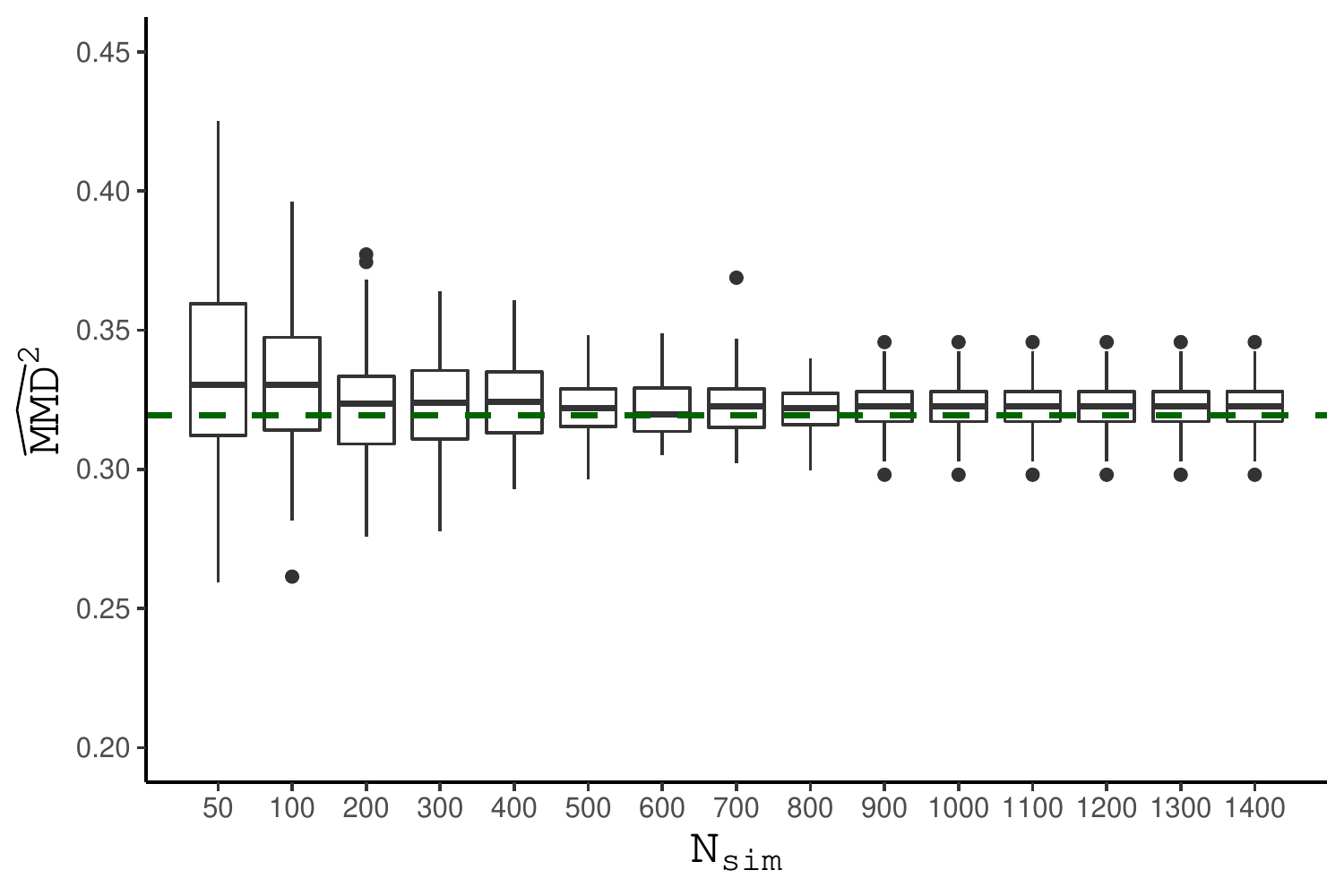}
  \caption{Boxplots of the estimated MMD$^2$ between $\mathbf{X}_{\mathrm{true}}$ ($\nreal = 1000$) and $\mathbf{X}'$ as a function of $\nsim$ computed by repeating the experiment 100 times for each value of $\nsim$. $\mathbf{X}'$ is generated from $\btheta' = [2\times 10^{-8}, 6\times 10^{7}, 10^{8}, 2\times 10^{-8},  10^{-9}, 5\times 10^{-10}]^\top$ and $\mathbf{X}_{\mathrm{true}}$ from $\btheta_{\mathrm{true}} = [5\times 10^{-8}, 2\times 10^{7}, 10^{9}, 10^{-8}, 2\times 10^{-9}, 10^{-9}]^\top$. The dashed green line is a corresponds to the value of the MMD$^2$ being approximated. Since this value is not available in closed-form, it is approximated by using $\nsim = 10^4$.}
      \label{fig:MMD_nsim}
\end{figure}

The seminal S-V model \cite{Saleh1987} is widely used as it is easy to simulate from, but is notoriously difficult to calibrate due to its structure. Even though the model can be analyzed using the theory of spatial point processes \cite{Jakobsen2012,Gubner2012} and moments derived \cite{Derpich2014}, its likelihood function is unavailable. Recent discussions of the physical interpretation of the S-V model, also outlining some difficulties with the model calibration, is given in \cite{Meijerinj2014, Pedersen2018, Pedersen2019}. These difficulties have inspired the use of many different  heuristic calibration methods, as outlined in the introduction. 

In the S-V model, the multipath components are assumed to arrive in clusters. The arrival time of the clusters and that of the rays within the clusters are modeled as one-dimensional homogeneous Poisson point processes with arrival rates $\Lambda$ and $\lambda$, respectively. The gains of the multipath components are modeled as iid zero-mean complex Gaussian random variables with conditional variance that depends on three parameters; the average power of the first arriving multipath component $Q$, and the cluster and ray power decay constants $\Gamma, \gamma$, respectively. We refer the readers to \cite{Saleh1987} and \cite{Jakobsen2012} for a detailed description of the model. Including the noise variance, the parameter vector becomes $\btheta = [Q, \Lambda, \lambda, \Gamma, \gamma, \sigma_W^2]^\top$.

We begin by finding a reasonable value of $\nsim$. To that end, we generate pseudo-observed log moments, $\mathbf{X}_{\mathrm{true}}$, with $\nreal = 1000$ realizations from the model by setting $\btheta$ to a ``true" value. Using another value of the parameter vector, say $\btheta'$, we simulate $\mathbf{X}'$ from the model with varying $\nsim$ and compute the estimated MMD between $\mathbf{X}'$ and $\textbf{X}_{\text{true}}$. This process is repeated $100$ times to create error bars as shown in Fig.~\ref{fig:MMD_nsim}. Although the MMD estimate gets more accurate as $\nsim$ increases, the improvement however is small. Choosing a higher $\nsim$ improves the MMD estimate, but increases the run-time of the of the algorithm significantly (since the computational cost is quadratic in $\nsim$, and simulating from the model can also be slow). Therefore, we set $\nsim = 100$ as a reasonable compromise considering the trade-off between accuracy and computational cost.

\begin{figure}
\centering
  \includegraphics[ width = 0.489\textwidth]{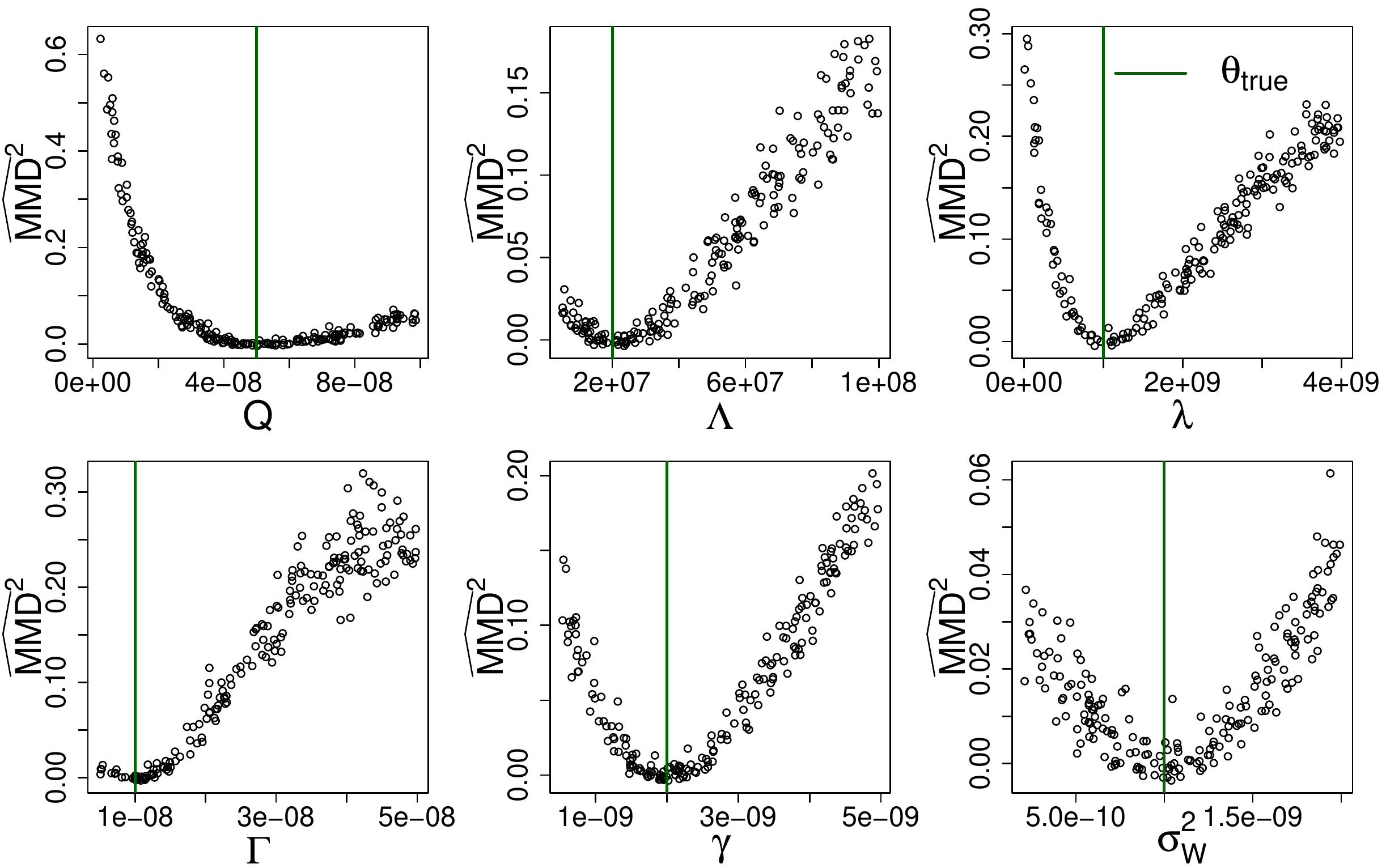}
  \caption{Estimated MMD$^2$ values plotted against parameters of the S-V model. The parameters are uniformly sampled 200 times from the prior range one at a time, keeping the others fixed to the true values denoted by the dark green lines. See Tab.~\ref{tab:estimates} for the prior ranges.}
      \label{fig:MMD_full}
\end{figure}

We first verify that the MMD computed from the temporal moments reacts to changes in the S-V model parameters. To that end, we generate simulated data-sets by varying one parameter uniformly in the prior support while keeping the others fixed to their true value. As can be seen from Fig.~\ref{fig:MMD_full}, the estimated MMD values increase as each of the parameters move away from their true value, and the minimum is (approximately) achieved when both the data-sets are generated from approximately the same parameters. The MMD reacts to changes in all the parameters, albeit more for some than others, as can be seen from the different scales of the y-axis. We therefore conclude that the distribution of the first four log temporal moments is informative about the S-V model parameters. 

\begin{figure}
\centering
  \includegraphics[ width = \columnwidth]{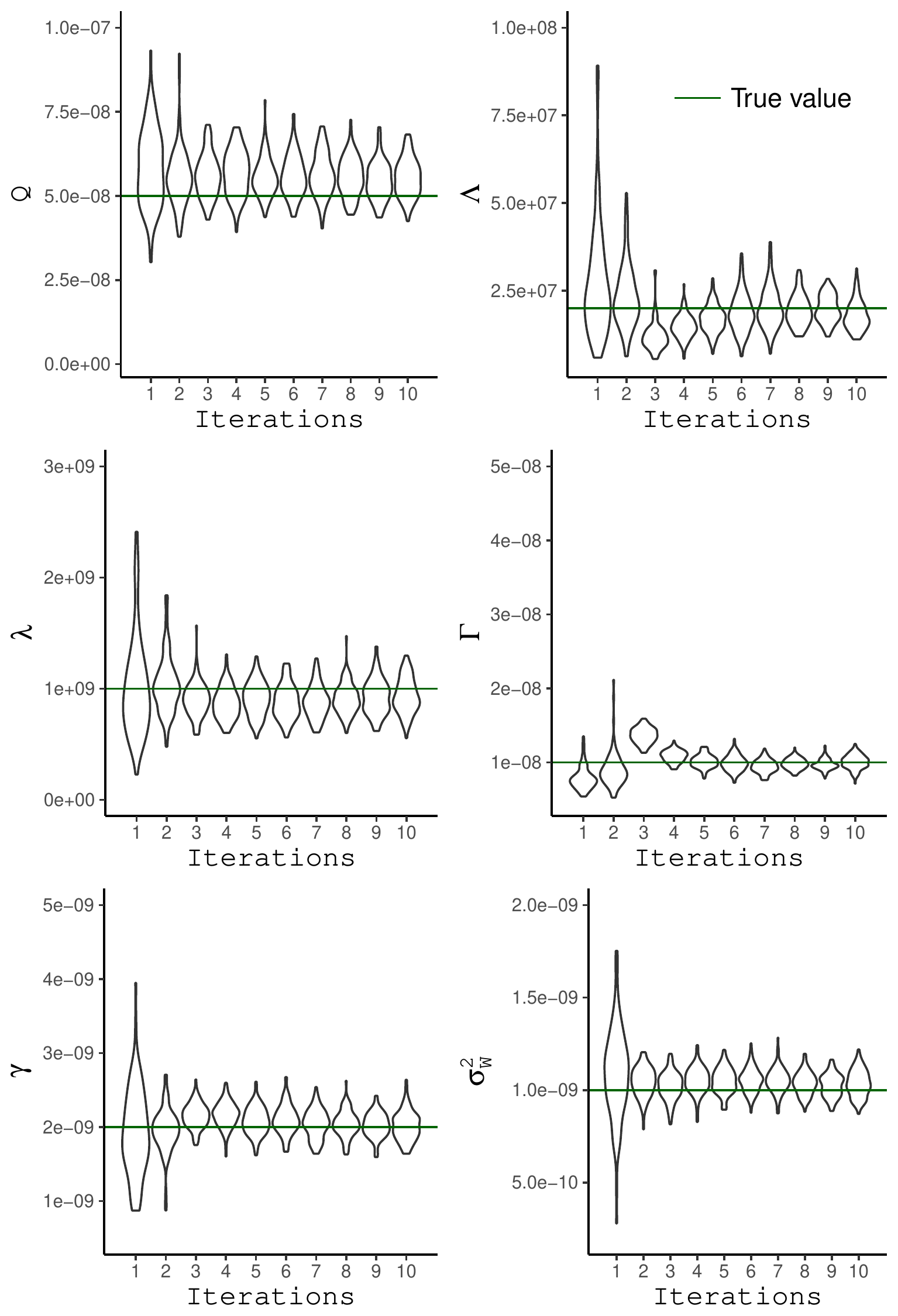}
  \caption{Violin plots of ABC posterior samples of S-V model parameters as a function of PMC iterations. Note that a violin plot is similar to a box plot with the addition of a rotated kernel density plot on each side. The dark green lines denote the true parameter values $\btheta_{\mathrm{true}} = [5\times 10^{-8}, 2\times 10^{7}, 10^{9}, 10^{-8}, 2\times 10^{-9}, 10^{-9}]^\top$.}
      \label{fig:box_SV_sim}
\end{figure}

We now use the proposed method to calibrate the S-V model using $\mathbf{X}_{\mathrm{true}}$. We assume uninformative (flat) priors in the range given in Tab.~\ref{tab:estimates} for all the parameters to ensure that their marginal posteriors are unaffected by any prior beliefs. The prior ranges were set according to the measurement settings as done in \cite{AyushABC}. The plots indicating convergence of the algorithm and the marginal posterior distributions for $T=10$ iterations are shown in Fig.~\ref{fig:box_SV_sim}. The approximate posterior samples concentrate around the true value for all the parameters. The algorithm converges rather quickly and the posteriors taper as the iterations proceed. In principle, the iterations could be stopped  after four or five iterations, but we let it run till $T=10$ for clarity. The algorithm gives a reasonable estimate for the parameters even in the first iteration. The proposed method is able to estimate $\Lambda$ accurately as well, unlike in \cite{AyushABC} where some post-processing was required to estimate $\Lambda$. 
\begin{table}[]
\centering
\caption{Parameter estimates obtained for measured data. The standard deviation of the approximate posterior samples is given in parenthesis.}
\resizebox{\columnwidth}{!}{%
\begin{tabular}{@{}cccc@{}}
\toprule
                           & $\btheta$          & \begin{tabular}[c]{@{}c@{}}\textbf{Prior range}\end{tabular} & \begin{tabular}[c]{@{}c@{}}\textbf{Estimate (std. deviation)}\end{tabular} \\ \midrule
\multirow{6}{*}{\rotatebox[origin=c]{90}{S-V model}} & $Q$                & $[10^{-9}, 10^{-7}]$                                   & $4.7 \times 10^{-8}$ ($4.6 \times 10^{-9}$)                                                               \\
                           & $\Lambda$          & $[5 \times 10^{6}, 10^{8}]$                            & $8.6 \times 10^{7}$ ($9.8 \times 10^{6}$)                                                                \\
                           & $\lambda$          & $[5 \times10^{-9}, 3 \times 10^{9}]$                   & $1.5 \times 10^{8}$ ($4.2 \times 10^{7}$)                                                               \\
                           & $\Gamma$           & $[5 \times 10^{-9}, 5 \times10^{-8}]$                  & $8.2 \times 10^{-9}$ ($2.7 \times 10^{-10}$)                                         \\
                           & $\gamma$           & $[5 \times 10^{-10}, 5 \times10^{-9}]$                 & $4.4 \times 10^{-9}$ ($4.7 \times 10^{-10}$)                                                               \\
                           & $\sigma_W^2$       & $[2 \times 10^{-10}, 2 \times10^{-9}]$                 & $3.5 \times 10^{-10}$ ($2.5 \times 10^{-11}$)                                                              \\ \midrule
\multirow{4}{*}{\rotatebox[origin=c]{90}{PG model}}  & $g$                & {[}0,1{]}                                              & 0.50 (0.019)                                                                            \\
                           & $N_{\mathrm{scat}}$ & {[}5,35{]}                                             & 18 (1.73)                                                                                \\
                           & $P_{\mathrm{vis}}$ & {[}0,1{]}                                              & 0.99 ($7.9 \times 10^{-4}$)                                                                              \\
                           & $\sigma_W^2$       & $[2 \times 10^{-10}, 2 \times10^{-9}]$                 & $4.4 \times 10^{-10}$ ($4.3 \times 10^{-12}$)                                                              \\ \bottomrule
\end{tabular}%
\label{tab:estimates}
}
\end{table}
\subsection{Application to the Propagation Graph model}
As our second example, we demonstrate the performance of our proposed method on the PG model. The PG model was first introduced in \cite{TPedersen2007}, and since then has been applied to a wide range of scenarios in \cite{LTian2012, LTian2016, JChenmmWave, Ramoni2019AWPL}. Recently, it has been extended to account for polarization in \cite{RamoniWCNC1,RamoniTAP2019,RamoniURSI}. Although the model is easy to simulate from, its likelihood function is unknown. A method of moments based estimator was applied to calibrate the model in \cite{RamoniTAP2019}, but the moments equations were based on approximation and it required manually fixing one of the parameters.

The PG model \cite{TPedersen2007} represents the radio channel as a directed graph with the transmitters, receivers and scatterers as vertices. Edges  model the wave propagation between the vertices. Edges are defined randomly depending on the probability of visibility, $P_{\mathrm{vis}}$. Other parameters of the model include the number of scatterers, $N_{\mathrm{scat}}$, and the reflection gain, $g$, resulting in the parameter vector $\btheta = [g, N_{\mathrm{scat}}, P_{\mathrm{vis}}, \sigma_W^2 ]^\top$. Note that $N_{\mathrm{scat}}$ is assumed to be real-valued during the regression adjustment, following which, its adjusted samples are rounded off to the nearest integer. We used the antenna positions and room geometry for the model according to the measurement conditions given in \cite{Gustafson2016}. Hence, $\nreal = \nsim = 625$ for the PG model. For each call of the model, the scatterer positions are drawn uniformly across the room, and all 625 realizations are generated based on those positions.

\begin{figure}
\centering
  \includegraphics[ width = \columnwidth]{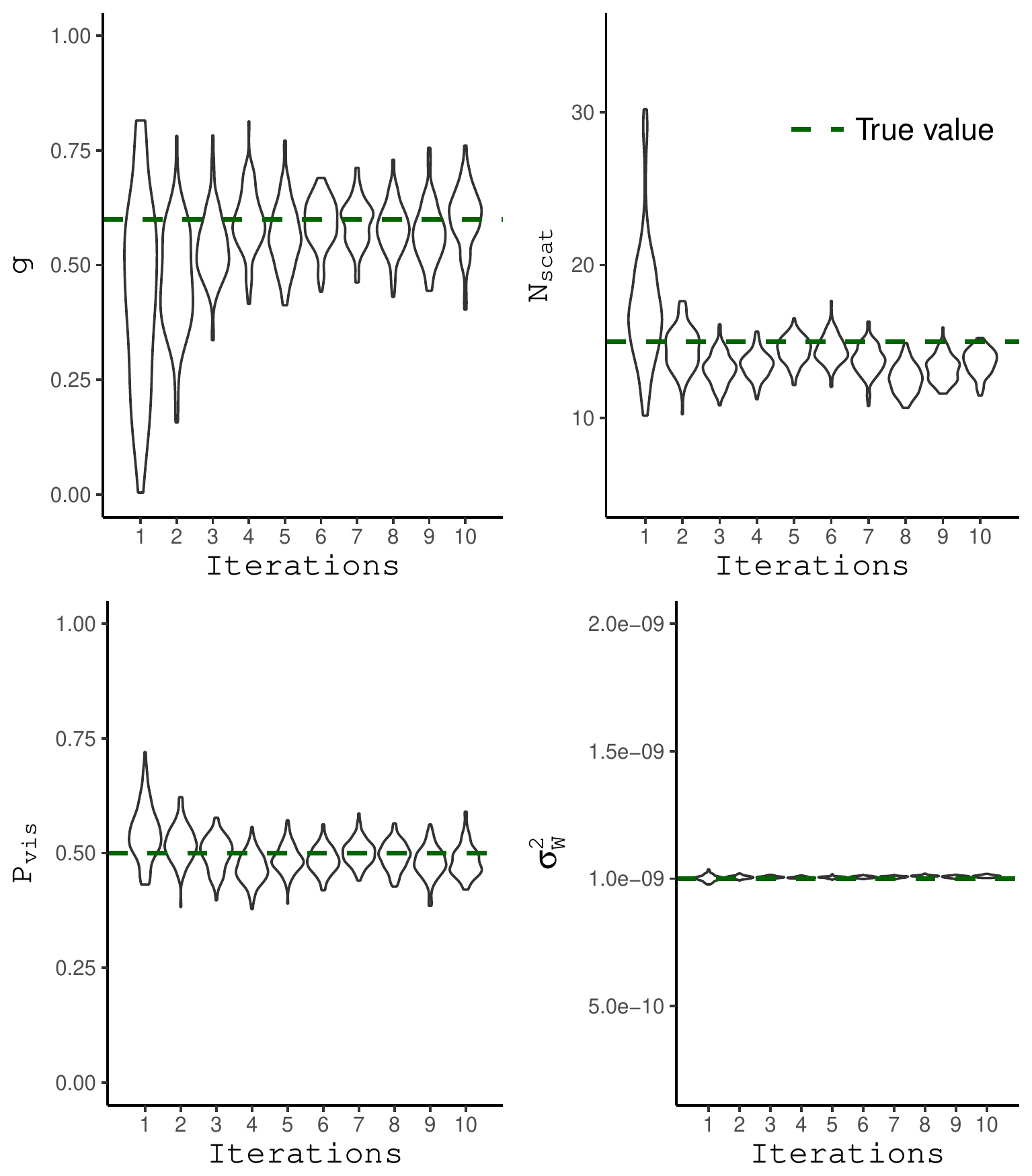}
  \caption{Violin plots of ABC posterior samples of PG model parameters as a function of PMC iterations. A violin plot is similar to a box plot with the addition of a rotated kernel density plot on each side. $\btheta_{\mathrm{true}} = [0.6, 15, 0.5, 10^{-9}]^\top$ is denoted by the dark green dashed line.} %The red line is $\hat{\btheta}_{\mathrm{MMSE}} = [0.54, 13, 0.5, 0.99 \times 10^{-9}]^\top$}
      \label{fig:box_pg_sim}
\end{figure}

We again use uniform priors for the parameters (see Tab.~\ref{tab:estimates}) and apply $T=10$ iterations of the proposed method to calibrate the PG model to the pseudo-observed data-set generated from $\btheta_{\mathrm{true}}$. To prevent biased results due to a particular configuration of the scatterers, we generate the pseudo-observed data by combining data from four different calls of the model using $\btheta_{\mathrm{true}}$. From Fig.~\ref{fig:box_pg_sim}, we observe that the algorithm converges very quickly, and gives posteriors which are highly concentrated around the true value for $P_{\mathrm{vis}}$, $N_{\mathrm{scat}}$, and $\sigma_W^2$. The approximate posterior for $g$ starts off very wide and then gets narrower as the iterations proceed. The method is therefore able to accurately calibrate the PG model. 

\begin{figure}
\centering
  \includegraphics[trim={60 220 75 230}, clip, width = \columnwidth]{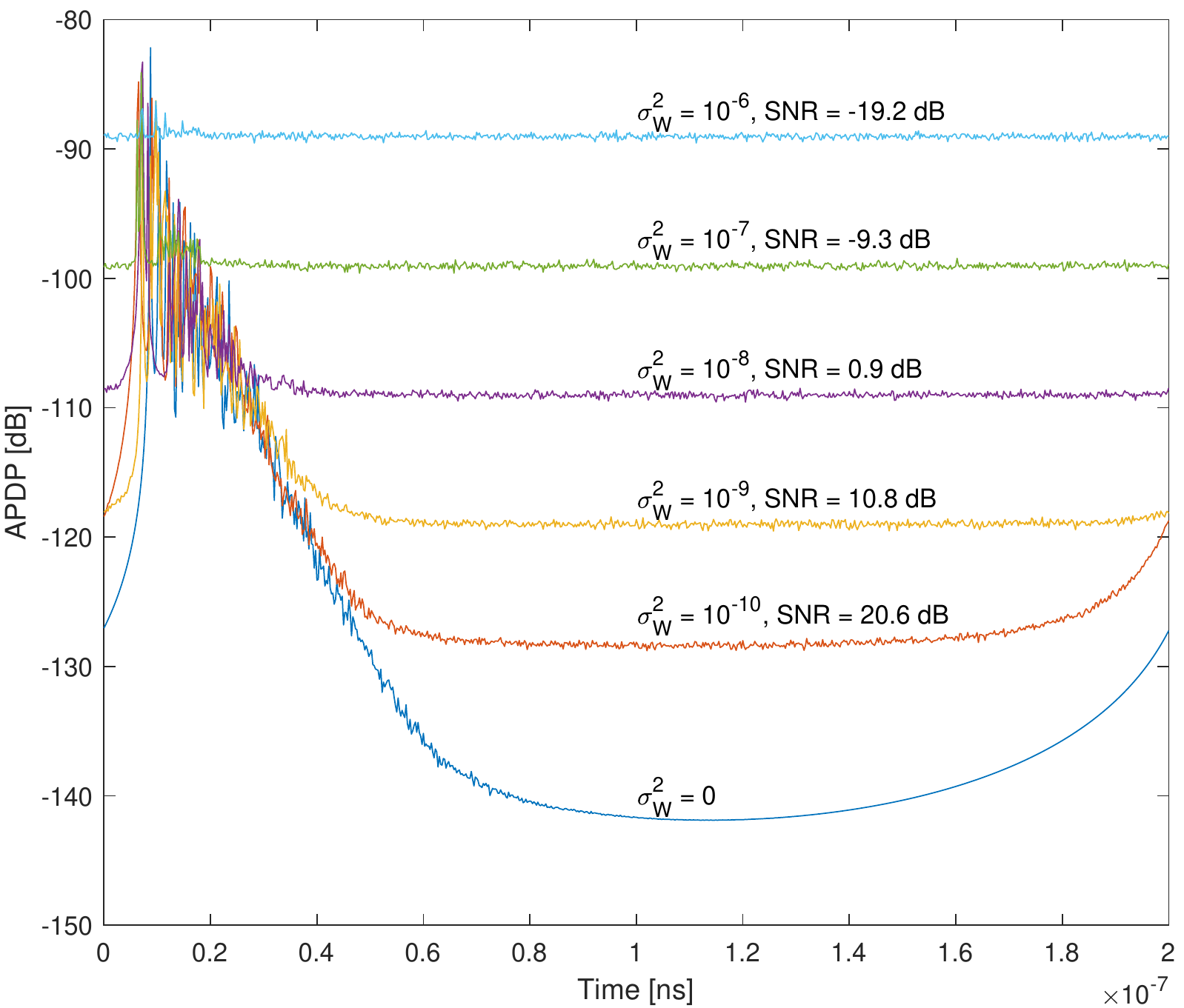}
  \caption{Averaged power delay profiles simulated from the PG model for different SNR levels.}
      \label{fig:snrAPDP}
\end{figure}

\begin{figure}
\centering
  \includegraphics[ width = \columnwidth]{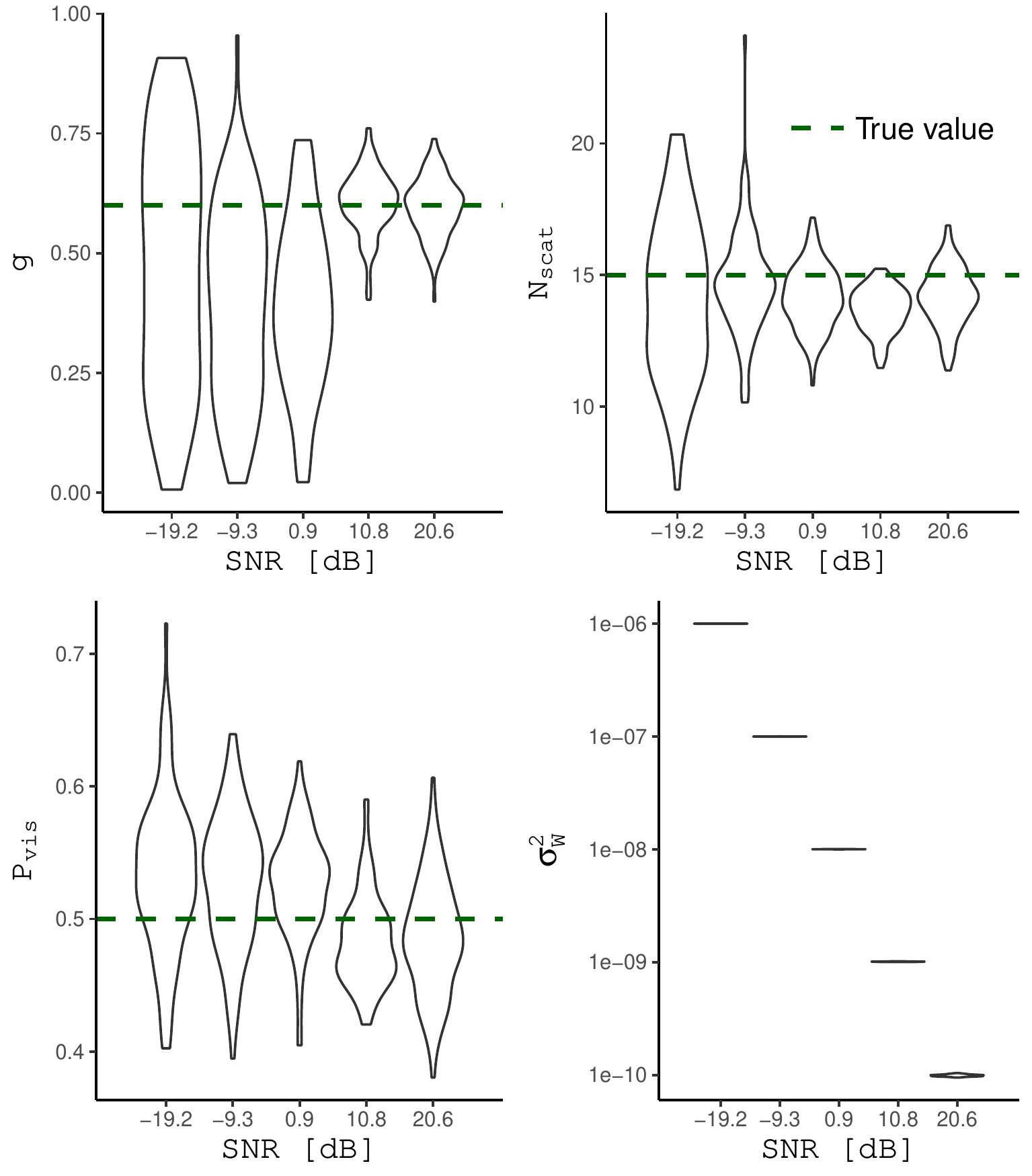}
  \caption{Violin plots of ABC posterior samples of PG model parameters after $T=10$ iterations for different SNR levels. The APDP corresponding to each SNR is shown in Fig.~\ref{fig:snrAPDP}.}
      \label{fig:snr}
\end{figure}

To asses how the performance of the proposed algorithm is affected by the presence of noise, we now repeat this simulation experiment for different noise levels. We fix $g=0.6$, $P_{\mathrm{vis}} = 0.5$, $N_{\mathrm{scat}} = 15$ and vary $\sigma^2_W$ from $10^{-10}$ to $10^{-6}$. The signal-to-noise ratio (SNR), is defined as
\begin{equation}
\text{SNR} = 10 \log_{10} (\bar m_0 B / \sigma^2_W) \quad [\text{dB}] ,    
\end{equation}
where $\bar{m}_0$ is the sample mean of of the zeroth temporal moment computed by setting $\sigma^2_W = 0$ in the PG model. The resulting averaged power delay profile (APDP) is shown in Fig.~\ref{fig:snrAPDP}. We run $T=10$ iterations of the algorithm for each of the SNR values. The prior for $\sigma^2_W$ is adjusted according to the true value in each run of the algorithm. The violin plots of the approximate posterior after the tenth iteration in each case is shown in Fig.~\ref{fig:snr}.

We observe that the noise variance $\sigma^2_W$ is estimated extremely accurately at each SNR level. The estimation accuracy for $P_{\mathrm{vis}}$ and $N_{\mathrm{scat}}$ seems to suffer only at the lowest SNR level. Reducing the SNR impacts the estimation accuracy of $g$ the most, with its approximate posterior converging to the prior as SNR decreases. This is expected as the higher the noise variance, the less visible the slope of the power delay profile which is determined by $g$. In conclusion, the algorithm performs well at SNR values encountered in measurements.

\section{Application to Measured Data}\label{sec:6}

We now attempt to fit both the S-V and the PG models to millimetre-wave radio channel measurements obtained from \cite{Gustafson2016}. The measurements of the channel transfer function were performed in the bandwidth 58 GHz to 62 GHz with a VNA, using $N_s = 801$ equally spaced frequency points. The bandwidth of $B = 4$ GHz means the frequency separation was $\Delta f = 5~\text{MHz}$ and $\tmax = 200~\text{ns}$. We use measurements taken in a small conference room of dimension $3\times 4 \times 3~\text{m}^3$ in a non-line-of-sight scenario. At both transmitter and receiver sides, $5\times5$ antenna arrays were used. Although the antenna elements used in the measurement were dual polarized, we focus on the vertical-vertical polarization since both the models are uni-polarized. This gives $\nreal = 5\times 5 \times 5 \times 5= 625$. We keep the settings $M = 2000$ and $M_\epsilon=100$ of the algorithm same as in the simulation experiments.

\subsection{Calibrating the Saleh-Valenzuela model}
\begin{figure}
\centering
  \includegraphics[ width = \columnwidth]{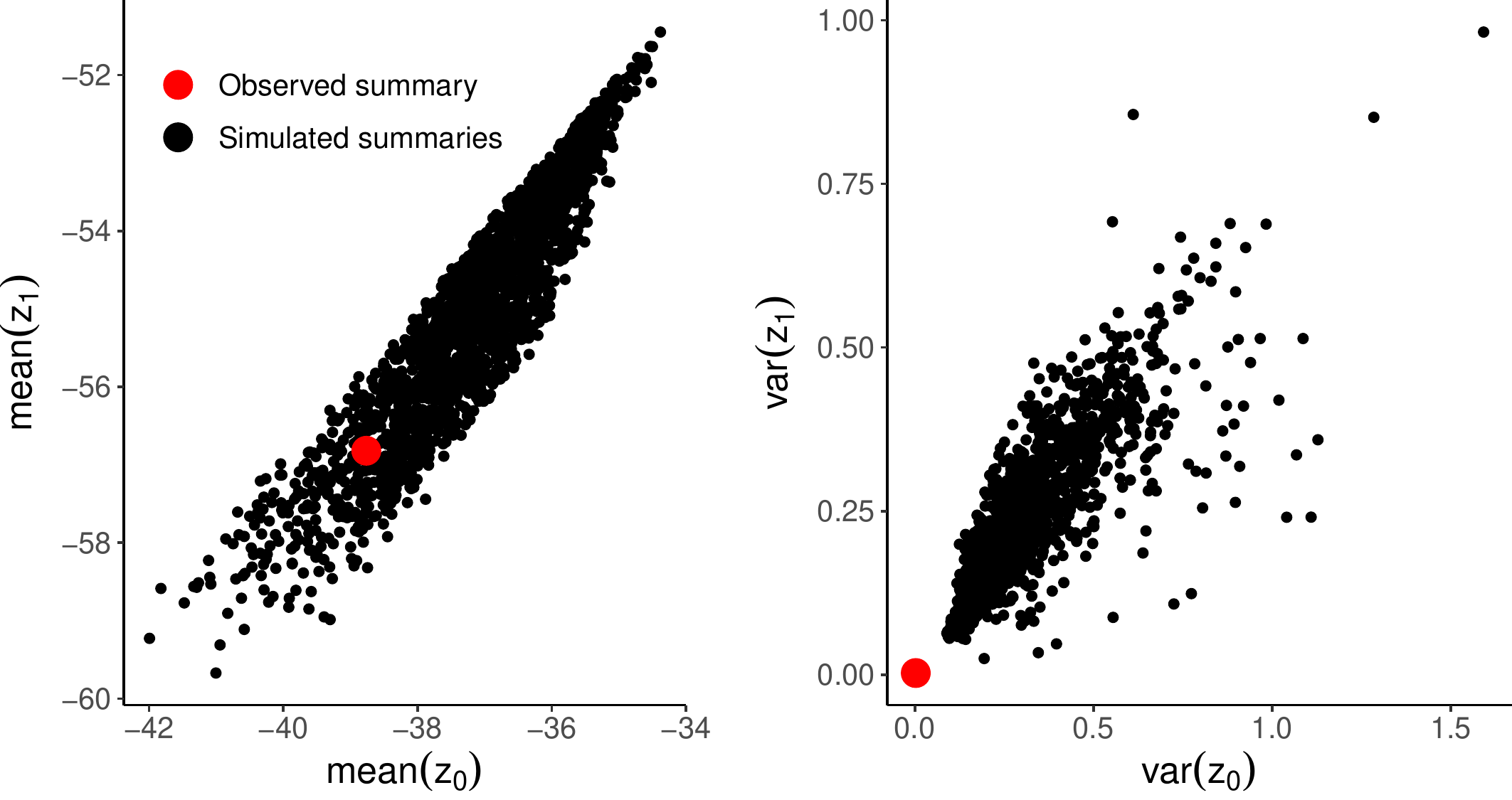}
  \caption{Mean (left) of $\mathbf{z}_0$ versus $\mathbf{z}_1$ simulated from the S-V model, along with the corresponding observed summary computed from the measured data (red). The mean and variance of $z_0$ is unitless, while that of $z_1$ is $[\text{log s}]$ and $[(\text{log s})^2]$, respectively. The observed summary lies in the point cloud generated by the model. In contrast, the S-V model is not able to replicate the higher moments of the data, as seen from the variance plot (right), indicating model misspecification. Each of the 2000 simulated summaries correspond to one parameter drawn from the prior.}
      \label{fig:scatter}
\end{figure}

Upon applying Alg.~\ref{alg:pmc-abc} to the measured data, regression adjustment yielded parameter samples outside the prior range. This indicated that the model is misspecified. That is indeed evident from Fig.~\ref{fig:scatter} where we plot elements of the vector $\mathbf{s}$, namely the mean and variance of $\mathbf{z}_0$ and $\mathbf{z}_1$, obtained from the measurements and the S-V model. The simulated summaries correspond to 2000 parameter values drawn from the prior. We observe that varying the parameters of the S-V model in the prior range generated mean values that overlap the mean value from the measurements. However, the variance values from the S-V model does not capture the value observed in the measurements. That is, there exists no such $\btheta$ in the prior range that leads to $\mathbf{s}$ ``close" to $\sobs$ in terms of the variance of the temporal moments. Hence, the model is misspecified for this data and so we obtain $\sobs$ from the model as per Sec.~\ref{sec:mismatch}. 

\begin{figure}
\centering
  \includegraphics[ width = \columnwidth]{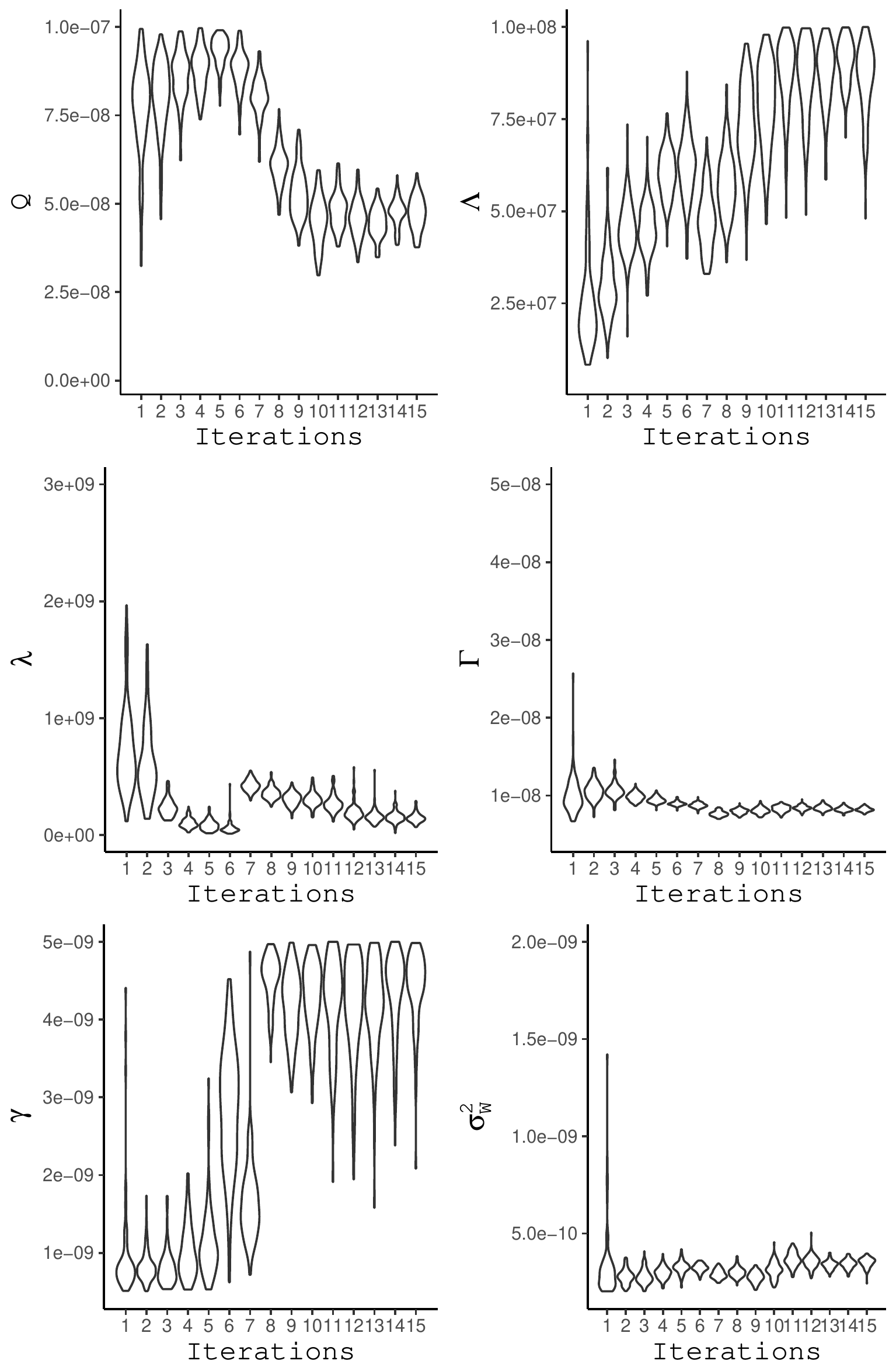}
  \caption{Violin plots of ABC posterior samples of S-V model parameters as a function of PMC iterations for measured data.}
      \label{fig:box_SV_meas}
\end{figure}

The posteriors obtained from the measured data are shown in Fig.~\ref{fig:box_SV_meas} for $T=15$ iterations. The marginal approximate posteriors for $\lambda$, $\Gamma$, and $\sigma^2_W$ are highly concentrated. Posteriors for $\Gamma$ and $\sigma^2_W$ appear to converge from the second iteration itself, indicating that these parameters affect the MMD the most. The posterior for $\lambda$ becomes narrow and converges after the first few iterations. The posteriors for $Q$, $\Lambda$ and $\gamma$ take around eight or nine iterations to converge to a different location in the prior range than where they began from, unlike the simulation experiment. This is potentially due to the model being misspecified for the data, and so parameters that affect the distribution of the log temporal moments the most converge first. The approximate estimates after 15 iterations are reported in Tab.~\ref{tab:estimates}. Considering that the regression adjustment in the first few iterations are done based on a coarse estimate of $\sobs$ from the model, the algorithm seems to work very well. The estimate of $\Lambda$ is high, indicating arrival of around 17 clusters on an average, while that of $\lambda$ is quite low. The model is therefore forced to the case with many clusters having very few multipath components each, thus approaching the ``unclustered" Turin model with constant rate.

The misspecification of the S-V model for the measured data is not surprising, as the measurement conditions are not replicated in the model. The virtual array measurements are from a single array  position in the room, hence the same clusters are observed in each transmit-receive antenna pair. On the other hand, each realization out of the S-V model is an independent realization from the underlying point process. As a result, we hardly see any variance in the log temporal moments of the data, which is not achieved in the S-V model for any configuration of the parameters.

\subsection{Calibrating the Propagation Graph model}
\begin{figure}
\centering
  \includegraphics[ width = \columnwidth]{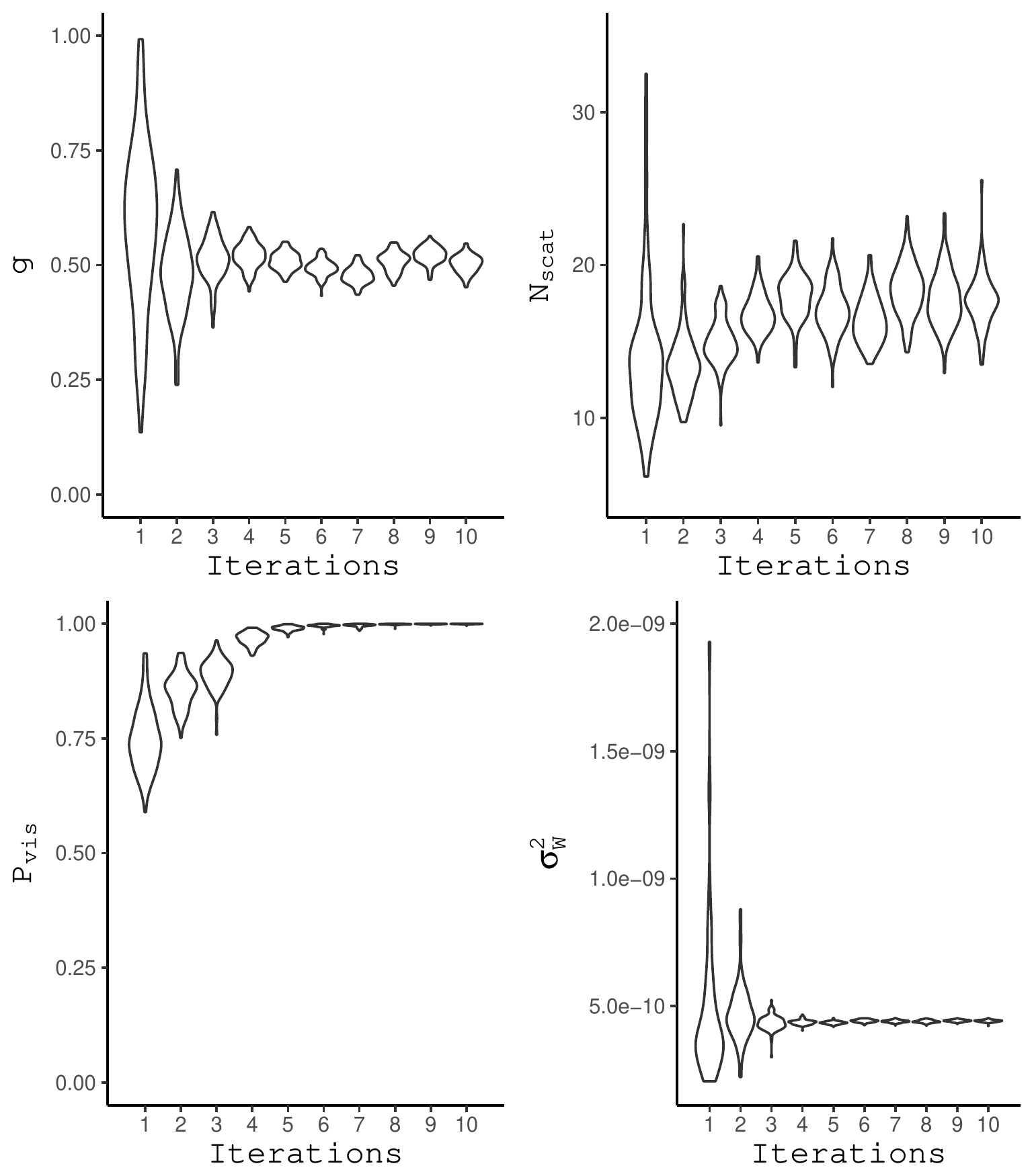}
  \caption{Violin plots of the ABC posterior samples of PG model parameters as a function of PMC iterations for  measured data.}
      \label{fig:box_pg_meas}
\end{figure}

The results obtained on calibration of the PG model on measured data after $T=10$ iterations is shown in Fig.~\ref{fig:box_pg_meas}. In this case, the model is not misspecified for the considered data. The approximate marginal posterior distributions for all the parameters start off wide and then seem to converge after around four or five iterations. The posteriors are also quite concentrated for all the four parameters, especially $P_{\mathrm{vis}}$ and $\sigma^2_W$. Overall, the results are similar to what is observed in the simulation experiment. See Tab.~\ref{tab:estimates} for approximate estimates of the parameters after $T=10$ iterations. The estimates are very similar to the ones reported in \cite{Bharti2020} where the polarized PG model was calibrated on data from the same measurement campaign. The estimate of $P_{\mathrm{vis}}$ is almost one, indicating that nearly all scatterers are connected. The estimates of $g$ and $P_{\mathrm{vis}}$ are consistent with the values reported from measurements \cite{RamoniTAP2019} in other in-room scenarios for the PG model. Moreover, these values are close to those used in simulations with the PG model in \cite{TPedersen2012,TPedersen2007}.

\subsection{Model Validation}
While the proposed method easily calibrates both the S-V and the PG models to measured data, there is no guarantee that the fitted models replicate the data well. This effect is of course not specific to the proposed method, but pertains to any calibration method. Thus, an extra step, termed model validation, should be performed where predictions of the calibrated models are compared to the data, and possibly other data-sets not used in the calibration process. Performing a full model validation is out of scope of this paper, as our focus is on the calibration method itself. Instead, as a final step we check how well the two calibrated models fit the input data-set.

\begin{figure}
\centering
  \includegraphics[ width = \columnwidth]{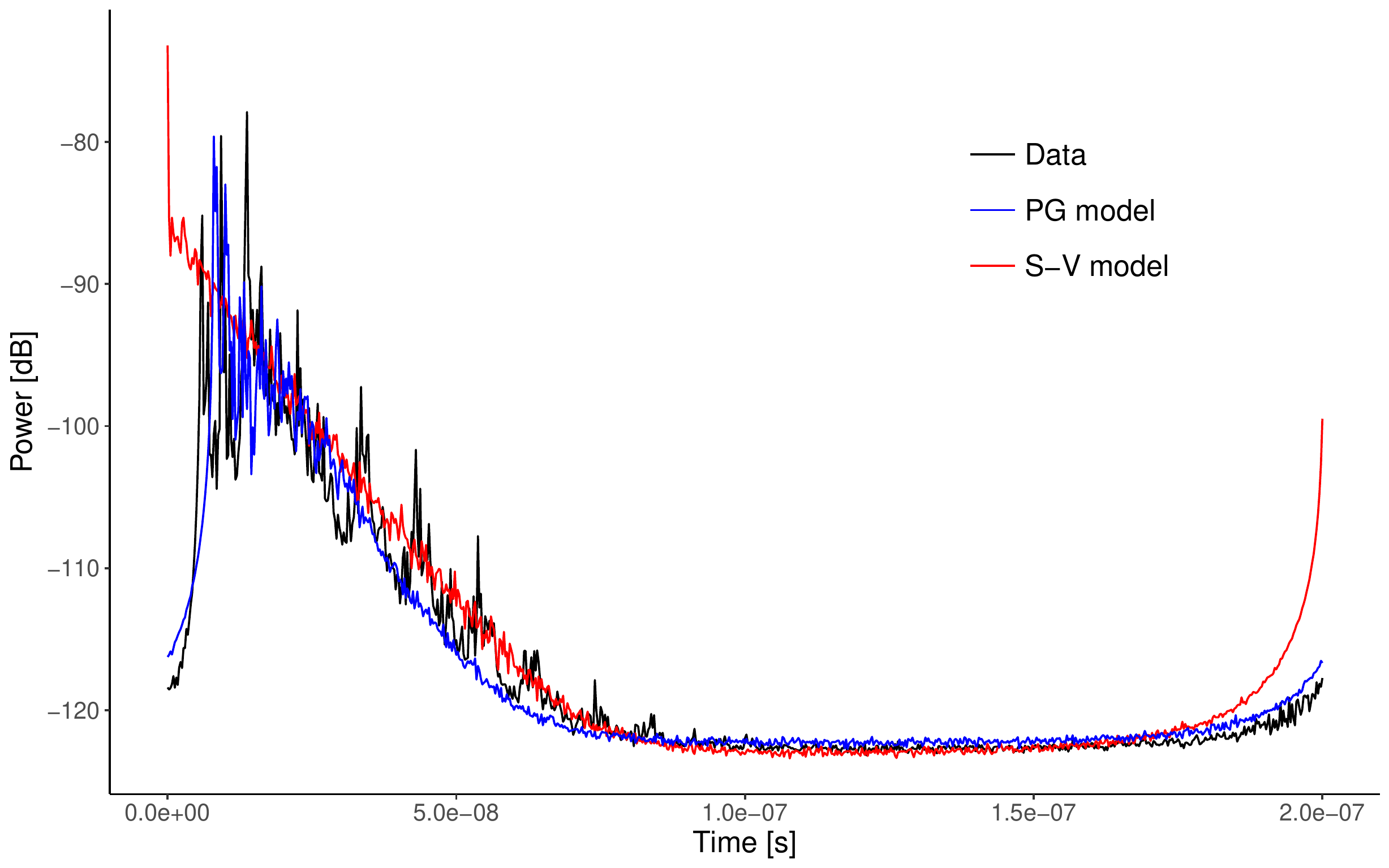}
  \includegraphics[ width = \columnwidth]{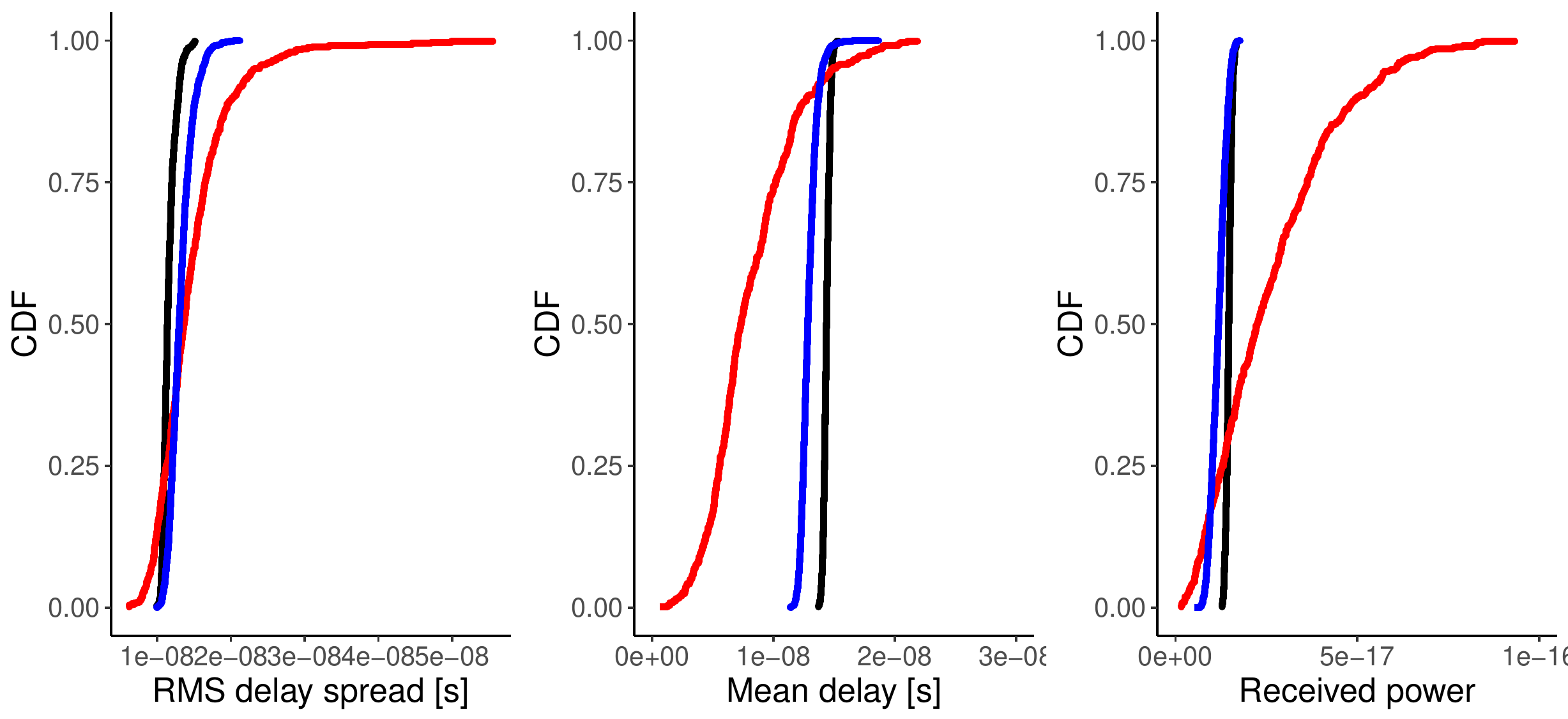}
  \caption{S-V and PG model fit to the measured data in terms of the APDP and empirical cdfs for rms delay spread, mean delay, and received power. Note that the first multipath component of the S-V model arrives at $t=0$ as in \cite{Saleh1987}.}
      \label{fig:apdp}
\end{figure}

To this end, we simulate 625 channel realizations from both models with parameters set according to Tab.~\ref{tab:estimates}. We compare the outputs from the models to the measured data in terms of the APDP and the empirical cumulative distribution function (cdf) of root mean square (rms) delay spread $\tau_{\mathrm{rms}}$, mean delay $\bar{\tau}$, and received power $P_0$ computed per channel realization, according to
\begin{equation}\label{eq:standardizedMoments}
P_0 = m_0, \quad \bar{\tau} = \frac{m_1}{m_0}, \quad \text{and} \quad \tau_{\mathrm{rms}} = \sqrt{\frac{m_2}{m_0} - \left(\frac{m_1}{m_0}\right)^2}.
\end{equation}
It appears from Fig.~\ref{fig:apdp}  that both the models are able to fit the APDP of the measurements well. The slope is captured well by both the models, along with the noise floor, although the S-V model slightly underestimates it. The S-V model, however, is not able to replicate the peaks in the APDP of the measurements, while the PG model represents the initial peaks better. This effect is to be expected for the particular settings of the S-V model with many clusters and very few within-cluster components. The peaks from the S-V model are averaged out since the channel realizations are independent. This is unlike the PG model where positions of the antennas in the virtual array are included, thus simulating correlated channel realizations.

Even though the APDPs are similar, the two models yields very different  empirical cdfs of $\tau_{\mathrm{rms}}$, $\bar{\tau}$, and $P_0$ as reported in Fig.~\ref{fig:apdp}. The PG model captures the behavior of the cdfs very well, while the S-V model clearly fails to do so, especially for the mean delay and the received power. The means of the rms delay spread from both the models are fairly close to the measured data, but the spread differs for the S-V model. As noticed theoretically in \cite{Pedersen2018,Pedersen2019}, multipath models can yield temporal moments with similar means while differing vastly in variance. Indeed, for a stochastic multipath model, the covariance structure of the temporal moments depends on both first- and second-order properties of the underlying point process \cite{Pedersen2020}.

The misspecification of the S-V model arises from disregarding the dependencies between the measurements obtained from different antennas in the array. This in turn leads to the discrepancy in the variance of the log temporal moments as observed in Fig~\ref{fig:scatter}. Thus, to alleviate the misspecification, the array structure should be incorporated in the model, as is done inherently in the PG model. This could be a contributing reason why other authors \cite{Salous2016} have found fully stochastic models inadequate and instead recommended using geometry-based and fully deterministic approaches for millimetre-wave data. Irrespective of the cause, such misspecifications can be detected by the proposed method, thereby assisting in the modeling process.

\section{Discussion}\label{sec:7}

The proposed method makes certain choices such as the number of temporal moments to use. We found that using the $I=4$ temporal moments, ($m_0,m_1,m_2,m_3$), gives accurate estimates with narrow posteriors after the first couple of iterations itself, while slight degradation in the performance is observed with $I=3$ moments $m_0$, $m_1$ and $m_2$. Although the method permits the use of arbitrarily many moments, we did not see significant improvements in performance when including more than four moments. Although the temporal moments seem adequate for calibration, the channel measurements could in principle be summarized into other statistics as long as they are informative about the model parameters.

Other choices for the method include the prior distribution and the settings of the ABC algorithm. We used uninformative priors to demonstrate the accuracy of the method based on data alone. However, including informative priors would speed up the convergence of the algorithm. For a reasonable approximation to the posterior distribution from samples, we suggest setting $M_\epsilon = 100$ or more. Depending on the computational budget, $\epsilon$ can be set around 5\% or less. Our chosen settings seem to work well for both the models, and hence, they can be a good starting point for initial experiments. We do not provide a stopping criterion for the algorithm, but instead encourage monitoring the posterior distributions for convergence, as the number of iterations required may vary across different parameters and models. Potentially a stopping criterion could be implemented where the iterations are stopped if the MMSE estimate changes less than some tolerance over iterations.

To calibrate a new channel model using our method, we suggest the following sequence of steps. Start by setting up priors for the model parameters based on available knowledge. Taking $J=4$ temporal moments as a starting point, perform the simulation study of computing the MMD$^2$ by varying one parameter at a time as done in Fig.~\ref{fig:MMD_full}. This experiment is informative in qualifying the required number of temporal moments. If the MMD is clearly impacted by varying the parameters, apply the method to calibrate the new model with the proposed settings of $M$ and $M_\epsilon$. If not, then adjust the number of temporal moments $J$ and repeat the process. Finally, monitor the posterior distributions for convergence and terminate the algorithm accordingly.

As the MMD compares infinitely many summaries of the two data-sets, it works better than comparing only the low-order moments such as the means and covariances of the temporal moments as in \cite{AyushSPAWC19, AyushABC, Bharti2020}. When choosing a characteristic kernel, the MMD also guarantees that distributions are uniquely identified by these moments, unlike the case when comparing a finite number of moments. The MMD is a strong notion of distance in the sense that recovery of the true parameter value is guaranteed as the number of data points grows. The MMD also leads to robust estimators; i.e. estimators which will return reasonable estimates even in the presence of outliers in the data or mild model misspecification \cite{Briol2019MMD,Cherief-Abdellatif2019finite}. The median heuristic is a reasonable choice for balancing robustness and efficiency as discussed in \cite{Briol2019MMD}. The choice of kernel is not as impactful as the choice of the lengthscale, and the proposed squared-exponential kernel seems to work well. 

The proposed method is computationally lightweight and can be run on standard laptops with reasonable run-time. In the experiments, the algorithm ran on a Lenovo ThinkPad with Intel Core i7 processor having 24~GB RAM. This gave a run-time of 5.5~hours for the PG model and around 2 days for the S-V model for ten iterations of the algorithm. In our tests, the computation time is dominated by the particular model evaluation time, while computation of temporal moments and the MMD is negligible. Thus, the computational cost depends heavily on the specific model and its implementation. Furthermore, the run-time is impacted by specific settings of some  parameters, e.g. $\Lambda$ and $\lambda$ in S-V model and $N_{\mathrm{scat}}$ in PG model. For higher ``true" values of these parameters, the model, and in turn, the calibration algorithm, takes considerably longer time to run. An obvious way to reduce the run-time is to run the algorithm on hardware with more processing power or by making parallel calls to the model during each iteration.

The proposed method relies solely on the ability to simulate from the model being calibrated, and not on the tractability of the likelihood or moment functions. Moreover, the method does not depend on the particular mathematical construction of the model, which enables calibration of very different models using the exact same procedure. This presents the opportunity to compare and select the best fitting model for a given data-set. Additionally, the proposed method inherently estimates the uncertainty of the fitted parameters, which is lacking in the state-of-the-art calibration approaches. In contrast to the rather complex state-of-the-art calibration methods, the proposed method is simple to implement in R, MATLAB or Python and requires very few settings such as $M$ and $M_\epsilon$. This has the clear advantage that results obtained from the method are easy to reproduce. Moreover, the method can detect and calibrate misspecified models as well. This is usually ignored or treated heuristically in standard algorithms. 

The proposed method can be used for a broad class of models where the likelihood is not known or difficult to compute. This is a great advantage in the model development as models can potentially be calibrated before their derivation is finalized. If the model is deemed worthy of further study, effort may be devoted to derive its likelihood function. The proposed method may also be used in  cases where such a likelihood is in fact available, or available up to some intractable normalization constant. In such cases the ABC approach may, however, be less effective than methods based on the likelihood. In those cases, other distances could be used; see for example Stein discrepancies for cases where the likelihood is unnormalised \cite{Barp2019}. Similarly, if factorization of the likelihood is possible and some factors can be evaluated, more efficient inference methods than ABC may be derived relying e.g. on message passing techniques. Such methods rely extensively on the particular models and the structure of their likelihoods. Thus, the gain in efficiency comes at a cost in the form of a loss in generality compared to the proposed ABC method. Finally, we remark that distance metrics such as the Wasserstein or the Hellinger distance could potentially be used instead of the MMD. However, future studies are required to assess their applicability for calibrating stochastic channel models. 

\section{Conclusion}\label{sec:8}

The proposed ABC method based on MMD is able to accurately calibrate wideband radio models of very different mathematical structure. The proposed method relies on computing temporal moments of the received signal, and thereby circumvents the need for multipath extraction or clustering. As a result, the method is automatic as no pre- or post-processing of the data and estimates are required. We find that the method is able to fit models to both simulated and measured data. This work opens possibilities of developing similar methods for calibrating directional and time-dependent channel models. Potentially, maximum mean discrepancy could be used for other problems in propagation and communication studies that involve comparing data-sets.

\section*{Acknowledgments}
The authors would like to thank Dr. Carl Gustafson and Prof. Fredrik Tufvesson (Lund University) for providing the measurement data.

\bibliography{bibliography}

\begin{thebibliography}{10}

\bibitem{Turin1972}
G.~L. Turin, F.~D. Clapp, T.~L. Johnston, S.~B. Fine, and D.~Lavry, ``A
  statistical model of urban multipath propagation,'' {\em IEEE Trans. Veh.
  Technol.}, vol.~21, pp.~1--9, Feb 1972.

\bibitem{Saleh1987}
A.~A.~M. Saleh and R.~Valenzuela, ``A statistical model for indoor multipath
  propagation,'' {\em IEEE J. Sel. Areas Commun.}, vol.~5, pp.~128--137,
  February 1987.

\bibitem{Haneda}
K.~Haneda, J.~J{\"a}rvel{\"a}inen, A.~Karttunen, M.~Kyr{\"o}, and J.~Putkonen,
  ``A statistical spatio-temporal radio channel model for large indoor
  environments at 60 and 70 {GH}z,'' {\em IEEE Trans. Antennas Propag.},
  vol.~63, no.~6, pp.~2694--2704, 2015.

\bibitem{METIS}
L.~Raschkowski, P.~Ky\"osti, K.~Kusume, and E.~T.~J\"ams\"a, {\em METIS channel
  models, deliverable D1.4 V3}.

\bibitem{WINNER}
P.~Ky\"osti, {\em WINNER II channel models, deliverables D1.1.2 V1.2, part I:
  Channel models}.

\bibitem{Gustafson}
C.~Gustafson, K.~Haneda, S.~Wyne, and F.~Tufvesson, ``On mm-wave multipath
  clustering and channel modeling,'' {\em IEEE Trans. Antennas Propag.},
  vol.~62, no.~3, pp.~1445--1455, 2014.

\bibitem{COST}
J.~Poutanen, K.~Haneda, L.~Liu, C.~Oestges, F.~Tufvesson, and P.~Vainikainen,
  ``Parameterization of the {COST} 2100 {MIMO} channel model in indoor
  scenarios,'' in {\em Eur. Conf. on Antennas and Propag.}, pp.~3606--3610,
  2011.

\bibitem{Li2019}
J.~{Li}, B.~{Ai}, R.~{He}, M.~{Yang}, Z.~{Zhong}, and Y.~{Hao}, ``A
  cluster-based channel model for massive mimo communications in indoor hotspot
  scenarios,'' {\em IEEE Trans. on Wireless Commun.}, vol.~18, no.~8,
  pp.~3856--3870, 2019.

\bibitem{Yang2020}
M.~{Yang}, B.~{Ai}, R.~{He}, G.~{Wang}, L.~{Chen}, X.~{Li}, C.~{Huang},
  Z.~{Ma}, Z.~{Zhong}, J.~{Wang}, Y.~{Li}, and T.~{Juhana}, ``Measurements and
  cluster-based modeling of vehicle-to-vehicle channels with large vehicle
  obstructions,'' {\em IEEE Trans. on Wireless Commun.}, vol.~19, no.~9,
  pp.~5860--5874, 2020.

\bibitem{Yin2018}
X.~Yin and X.~Cheng, {\em Propagation Channel Characterization, Parameter
  Estimation, and Modeling for Wireless Communications}.
\newblock John Wiley {\&} Sons Singapore Pte. Ltd, Feb 2018.

\bibitem{Czink}
N.~Czink, P.~Cera, J.~Salo, E.~Bonek, J.~P. Nuutinen, and J.~Ylitalo, ``A
  framework for automatic clustering of parameteric {MIMO} channel data
  including path powers,'' in {\em Proc. IEEE 64th Veh. Technol. Conf.-Fall},
  pp.~1--5, 2006.

\bibitem{Gentile}
C.~Gentile, ``Using the kurtosis measure to identify clusters in wireless
  channel impulse responses,'' {\em IEEE Trans. Antennas Propag.}, vol.~61,
  no.~6, pp.~3392--3396, 2013.

\bibitem{Ruisi}
R.~He, W.~Chen, B.~Ai, A.~F. Molisch, W.~Wang, Z.~Zhong, J.~Yu, and
  S.~Sangodoyin, ``On the clustering of radio channel impulse responses using
  sparsity-based methods,'' {\em IEEE Trans. Antennas Propag.}, vol.~64, no.~6,
  pp.~2465--2474, 2016.

\bibitem{Greenstein2007}
L.~Greenstein, S.~Ghassemzadeh, S.-C. Hong, and V.~Tarokh, ``Comparison study
  of {UWB} indoor channel models,'' {\em {IEEE} Trans. on Wireless Commun.},
  vol.~6, pp.~128--135, Jan 2007.

\bibitem{RamoniTAP2019}
R.~Adeogun, T.~Pedersen, C.~Gustafson, and F.~Tufvesson, ``{Polarimetric
  Wireless Indoor Channel Modelling Based on Propagation Graph},'' {\em IEEE
  Trans. on Antennas and Propag.}, vol.~67, no.~10, pp.~6585--6595, 2019.

\bibitem{Hirsch2020}
C.~Hirsch, A.~Bharti, T.~Pedersen, and R.~Waagepetersen, ``Maximum likelihood
  calibration of stochastic multipath radio channel models,'' {\em {IEEE}
  Trans. on Antennas and Propag.}, pp.~1--1, 2020.

\bibitem{AyushURSI}
A.~Bharti, R.~Adeogun, and T.~Pedersen, ``{Parameter Estimation for Stochastic
  Channel Models using Temporal Moments},'' in {\em Proc. 2019 IEEE Int. Symp.
  on Antennas and Propag. and USNC-URSI Radio Sci. Meeting}, pp.~1267--1268,
  2019.

\bibitem{Wu2008}
W.-D. Wu, C.-H. Wang, C.-C. Chao, and K.~Witrisal, ``On parameter estimation
  for ultra-wideband channels with clustering phenomenon,'' in {\em {IEEE} 68th
  Veh. Technol. Conf.}, {IEEE}, Sep 2008.

\bibitem{AyushSPAWC19}
A.~Bharti, R.~Adeogun, and T.~Pedersen, ``{Estimator for Stochastic Channel
  Model without Multipath Extraction using Temporal Moments},'' in {\em 20th
  IEEE Int. Workshop on Signal Process. Advances in Wireless Commun. (SPAWC)},
  pp.~1--5, 2019.

\bibitem{AyushABC}
A.~Bharti and T.~Pedersen, ``Calibration of stochastic channel models using
  approximate {B}ayesian computation,'' in {\em Proc. IEEE Global Commun. Conf.
  Workshops}, pp.~1--6, 2019.

\bibitem{RamoniMLAWPL}
R.~{Adeogun}, ``Calibration of stochastic radio propagation models using
  machine learning,'' {\em IEEE Antennas and Wireless Propag. Lett.}, vol.~18,
  pp.~2538--2542, Dec 2019.

\bibitem{Bharti2020}
A.~{Bharti}, R.~{Adeogun}, and T.~{Pedersen}, ``Learning parameters of
  stochastic radio channel models from summaries,'' {\em IEEE Open J. of
  Antennas and Propag.}, vol.~1, pp.~175--188, 2020.

\bibitem{AyushEuCAP21}
A.~Bharti, R.~Adeogun, and T.~Pedersen, ``Auto-generated summaries for
  stochastic radio channel models,'' in {\em 15th Eur. Conf. on Antennas and
  Propag.}, pp.~1--5, 2021.

\bibitem{Sisson2018}
S.~A. Sisson, {\em Handbook of Approximate Bayesian Computation}.
\newblock Chapman and Hall/{CRC}, Sep 2018.

\bibitem{Gretton2012JMLR}
A.~Gretton, K.~Borgwardt, M.~J. Rasch, and B.~Scholkopf, ``{A kernel two-sample
  test},'' {\em J. of Mach. Learn. Res.}, vol.~13, pp.~723--773, 2012.

\bibitem{Briol2019MMD}
F.-X. Briol, A.~Barp, A.~B. Duncan, and M.~Girolami, ``{Statistical inference
  for generative models with maximum mean discrepancy},'' {\em
  arXiv:1906.05944}, 2019.

\bibitem{Cherief-Abdellatif2019finite}
B.-E. Ch{\'{e}}rief-Abdellatif and P.~Alquier, ``{Finite sample properties of
  parametric MMD estimation: robustness to misspecification and dependence},''
  {\em arXiv:1912.05737}, 2020.

\bibitem{Cherief-Abdellatif2019MMDBayes}
B.-E. Cherief-Abdellatif and P.~Alquier, ``{MMD-Bayes}: Robust bayesian
  estimation via maximum mean discrepancy,'' vol.~118 of {\em Proc. of Mach.
  Learn. Res.}, pp.~1--21, PMLR, 08 Dec 2020.

\bibitem{Nakagome2013}
S.~Nakagome, K.~Fukumizu, and S.~Mano, ``{Kernel approximate Bayesian
  computation in population genetic inferences},'' {\em Stat. Appl. in Genet.
  and Mol. Biol.}, vol.~12, no.~6, pp.~667--678, 2013.

\bibitem{Park2015}
M.~Park, W.~Jitkrittum, and D.~Sejdinovic, ``{K2-ABC: approximate Bayesian
  computation with kernel embeddings},'' {\em Proc. of the 19th Int. Conf. on
  Artif. Intell. and Statistics}, vol.~51, pp.~398--407, 2015.

\bibitem{Mitrovic2016}
J.~Mitrovic, D.~Sejdinovic, and Y.~W. Teh, ``{DR-ABC: Approximate Bayesian
  computation with kernel-based distribution regression},'' {\em 33rd Int.
  Conf. on Mach. Learn., ICML}, vol.~3, pp.~2209--2218, 2016.

\bibitem{Kisamori2020}
K.~Kisamori, M.~Kanagawa, and K.~Yamazaki, ``Simulator calibration under
  covariate shift with kernels,'' in {\em Proc.s of the 23rd Int. Conf. on
  Artif. Intell. and Statistics}, vol.~108, pp.~1244--1253, PMLR, Aug 2020.

\bibitem{Dziugaite2015}
G.~K. Dziugaite, D.~M. Roy, and Z.~Ghahramani, ``{Training generative neural
  networks via maximum mean discrepancy optimization},'' in {\em Proc. of 31st
  Conf. on Uncertain. in Artif. Intell.}, pp.~258--267, 2015.

\bibitem{Sutherland2017}
D.~J. Sutherland, H.-Y. Tung, H.~Strathmann, S.~De, A.~Ramdas, A.~Smola, and
  A.~Gretton, ``{Generative models and model criticism via optimized maximum
  mean discrepancy},'' in {\em Int. Conf. on Learn. Represent.}, 2017.

\bibitem{Li2015GMMN}
Y.~Li, K.~Swersky, and R.~Zemel, ``Generative moment matching networks,'' in
  {\em Proc. of the 32nd Int. Conf. on Mach. Learn. - Vol. 37}, ICML'15,
  p.~1718–1727, JMLR.org, 2015.

\bibitem{Muandet2017}
K.~Muandet, K.~Fukumizu, B.~Sriperumbudur, and B.~Schölkopf, ``Kernel mean
  embedding of distributions: A review and beyond,'' {\em Found. and Trends in
  Mach. Learn.}, vol.~10, no.~1-2, pp.~1--141, 2017.

\bibitem{Berlinet2004}
A.~Berlinet and C.~Thomas-Agnan, {\em {Reproducing Kernel Hilbert Spaces in
  Probability and Statistics}}.
\newblock New York: Springer Science+Business Media, 2004.

\bibitem{Sriperumbudur2009}
B.~K. Sriperumbudur, A.~Gretton, K.~Fukumizu, B.~Sch{\"{o}}lkopf, and G.~R.~G.
  Lanckriet, ``{Hilbert space embeddings and metrics on probability
  measures},'' {\em J. of Mach. Learn. Res.}, vol.~11, 2010.

\bibitem{Simon-Gabriel2016}
C.-J. Simon-Gabriel and B.~Sch{\"{o}}lkopf, ``{Kernel Distribution Embeddings:
  Universal Kernels, Characteristic Kernels and Kernel Metrics on
  Distributions},'' {\em J. of Mach. Learn. Res.}, vol.~19, no.~44, pp.~1--29,
  2018.

\bibitem{Briol2015}
F.-X. Briol, C.~J. Oates, M.~Girolami, and M.~A. Osborne, ``Frank-wolfe
  bayesian quadrature: Probabilistic integration with theoretical guarantees,''
  in {\em Proc. of the 28th Int. Conf. on Neural Inf. Process. Syst. - Volume
  1}, NIPS'15, p.~1162–1170, 2015.

\bibitem{Reddi2015}
S.~Reddi, A.~Ramdas, B.~Poczos, A.~Singh, and L.~Wasserman, ``{On the High
  Dimensional Power of a Linear-Time Two Sample Test under Mean-shift
  Alternatives},'' in {\em Proc. of the 18th Int. Conf. on Artif. Intell. and
  Stat.}, vol.~38, pp.~772--780, PMLR, May 2015.

\bibitem{Ruping2001}
S.~R{\"{u}}ping, ``{SVM kernels for time series analysis},'' tech. rep., 2001.

\bibitem{Cuturi2007}
M.~Cuturi, J.-P. Vert, {\O}.~Birkenes, and T.~Matsui, ``A kernel for time
  series based on global alignments,'' {\em IEEE Int. Conf. on Acoust,, Speech
  and Signal Process.}, vol.~2, pp.~413--416, 2007.

\bibitem{Cuturi2011A}
M.~Cuturi, ``{Fast global alignment kernels},'' {\em Proc. of the 28th Int.
  Conf. on Mach. Learn.}, pp.~929--936, 2011.

\bibitem{Chevyrev2018}
I.~Chevyrev and H.~Oberhauser, ``{Signature moments to characterize laws of
  stochastic processes},'' {\em arXiv:1810.10971}, 2018.

\bibitem{Kiraly2019}
F.~J. Kiŕaly and H.~Oberhauser, ``{Kernels for sequentially ordered data},''
  {\em J. of Mach. Learn. Res.}, vol.~20, pp.~1--45, 2019.

\bibitem{Wynne2020}
G.~Wynne and A.~B. Duncan, ``A kernel two-sample test for functional data,''
  {\em arXiv:2008.11095}, 2020.

\bibitem{Franks1969}
L.~E. Franks, {\em {Signal Theory}}.
\newblock Englewood Cliffs, N. J., Prentice-Hall, 1969.

\bibitem{AyushEuCAP}
A.~Bharti, L.~Clavier, and T.~Pedersen, ``Joint statistical modeling of
  received power, mean delay, and delay spread for indoor wideband radio
  channels,'' in {\em 14th Eur. Conf. on Antennas and Propag.}, pp.~1--5, 2020.

\bibitem{Bharti2021}
A.~Bharti, R.~Adeogun, X.~Cai, W.~Fan, F.-X. Briol, L.~Clavier, and
  T.~Pedersen, ``Joint modeling of received power, mean delay, and delay spread
  for wideband radio channels,'' {\em {IEEE} Trans. on Antennas and Propag.},
  pp.~1--1, 2021.

\bibitem{Steinwart2008}
I.~Steinwart and A.~Christmann, {\em {Support Vector Machines}}.
\newblock Springer, 2008.

\bibitem{Beaumont2009}
M.~A. Beaumont, J.-M. Cornuet, J.-M. Marin, and C.~P. Robert, ``Adaptive
  approximate bayesian computation,'' {\em Biometrika}, vol.~96, pp.~983--990,
  Oct 2009.

\bibitem{Beaumont2002}
M.~A. Beaumont, W.~Zhang, and D.~J. Balding, ``Approximate bayesian computation
  in population genetics,'' {\em Genetics}, vol.~162, no.~4, pp.~2025--2035,
  2002.

\bibitem{Lintusaari2016}
J.~Lintusaari, M.~U. Gutmann, R.~Dutta, S.~Kaski, and J.~Corander,
  ``Fundamentals and recent developments in approximate bayesian computation,''
  {\em Syst. Biol.}, vol.~66, pp.~66--82, Jan 2017.

\bibitem{Frazier2020}
D.~T. Frazier, C.~P. Robert, and J.~Rousseau, ``Model misspecification in
  approximate bayesian computation: consequences and diagnostics,'' {\em J. of
  the R. Stat. Soc.: Ser. B (Stat. Methodol.)}, vol.~82, pp.~421--444, Jan
  2020.

\bibitem{Gustafson2016}
C.~Gustafson, D.~Bolin, and F.~Tufvesson, ``Modeling the polarimetric mm-wave
  propagation channel using censored measurements,'' in {\em 2016 Global
  Commun. Conf.}, {IEEE}, Dec 2016.

\bibitem{Jakobsen2012}
M.~L. Jakobsen, T.~Pedersen, and B.~H. Fleury, ``Analysis of the stochastic
  channel model by saleh \& valenzuela via the theory of point processes,''
  {\em Int. Zurich Seminar on Commun.}, 2012.

\bibitem{Gubner2012}
J.~A. {Gubner}, B.~N. {Bhaskar}, and K.~{Hao}, ``Multipath-cluster channel
  models,'' in {\em 2012 IEEE Int. Conf. on Ultra-Wideband}, pp.~292--296,
  2012.

\bibitem{Derpich2014}
M.~S. {Derpich} and R.~{Feick}, ``Second-order spectral statistics for the
  power gain of wideband wireless channels,'' {\em IEEE Trans. on Veh.
  Technol.}, vol.~63, no.~3, pp.~1013--1031, 2014.

\bibitem{Meijerinj2014}
A.~{Meijerink} and A.~F. {Molisch}, ``On the physical interpretation of the
  saleh–valenzuela model and the definition of its power delay profiles,''
  {\em IEEE Trans. on Antennas and Propag.}, vol.~62, no.~9, pp.~4780--4793,
  2014.

\bibitem{Pedersen2018}
T.~{Pedersen}, ``Modeling of path arrival rate for in-room radio channels with
  directive antennas,'' {\em IEEE Trans. on Antennas and Propag.}, vol.~66,
  no.~9, pp.~4791--4805, 2018.

\bibitem{Pedersen2019}
T.~{Pedersen}, ``{Stochastic Multipath Model for the In-Room Radio Channel
  Based on Room Electromagnetics},'' {\em IEEE Trans. on Antennas and Propag.},
  vol.~67, pp.~2591--2603, April 2019.

\bibitem{TPedersen2007}
T.~Pedersen and B.~H. Fleury, ``Radio channel modelling using stochastic
  propagation graphs,'' in {\em IEEE ICC}, pp.~2733--2738, June 2007.

\bibitem{LTian2012}
L.~Tian, X.~Yin, Q.~Zuo, J.~Zhou, Z.~Zhong, and S.~X. Lu, ``Channel modeling
  based on random propagation graphs for high speed railway scenarios,'' in
  {\em IEEE PIMRC}, pp.~1746--1750, Sept 2012.

\bibitem{LTian2016}
L.~Tian, V.~Degli-Esposti, E.~M. Vitucci, and X.~Yin, ``Semi-deterministic
  radio channel modeling based on graph theory and ray-tracing,'' {\em IEEE
  Trans. on Antennas and Propag.}, vol.~64, pp.~2475--2486, June 2016.

\bibitem{JChenmmWave}
J.~{Chen}, X.~{Yin}, L.~{Tian}, and M.~{Kim}, ``Millimeter-wave channel
  modeling based on a unified propagation graph theory,'' {\em IEEE Commun.
  Lett.}, vol.~21, pp.~246--249, Feb 2017.

\bibitem{Ramoni2019AWPL}
R.~O. {Adeogun}, A.~{Bharti}, and T.~{Pedersen}, ``An iterative transfer matrix
  computation method for propagation graphs in multi-room environments,'' {\em
  IEEE Antennas and Wireless Propag. Lett.}, vol.~18, pp.~616--620, April 2019.

\bibitem{RamoniWCNC1}
R.~Adeogun and T.~Pedersen, ``Propagation graph based model for multipolarized
  wireless channels,'' in {\em {IEEE WCNC}}, April 2018.

\bibitem{RamoniURSI}
R.~Adeogun and T.~Pedersen, ``Modelling polarimetric power delay spectrum for
  indoor wireless channels via propagation graph formalism,'' in {\em {2nd URSI
  Atlantic Radio Sci. Meeting}}, May 2018.

\bibitem{TPedersen2012}
T.~Pedersen, G.~Steinb{\"o}ck, and B.~H. Fleury, ``Modeling of reverberant
  radio channels using propagation graphs,'' vol.~60, pp.~5978--5988, Dec 2012.

\bibitem{Pedersen2020}
T.~{Pedersen}, ``First- and second order characterization of temporal moments
  of stochastic multipath channels,'' in {\em 2020 33rd Gen. Assembly and Sci.
  Symp. of the Int. Union of Radio Sci.}, pp.~1--4, 2020.

\bibitem{Salous2016}
S.~Salous, V.~D. Esposti, F.~Fuschini, R.~S. Thomae, R.~Mueller, D.~Dupleich,
  K.~Haneda, J.-M.~M. Garcia-Pardo, J.~P. Garcia, D.~P. Gaillot, S.~Hur, and
  M.~Nekovee, ``Millimeter-wave propagation: Characterization and modeling
  toward fifth-generation systems. [wireless corner],'' {\em {IEEE} Antennas
  and Propagation Magazine}, vol.~58, pp.~115--127, dec 2016.

\bibitem{Barp2019}
A.~Barp, F.-X. Briol, A.~Duncan, M.~Girolami, and L.~Mackey, ``Minimum stein
  discrepancy estimators,'' in {\em Adv. in Neural Inform. Process. Syst.},
  vol.~32, pp.~12964--12976, 2019.

\end{thebibliography}
\bibliographystyle{ieeetr}

\begin{IEEEbiography}[{\includegraphics[width=1in,height=1.25in,clip,keepaspectratio]{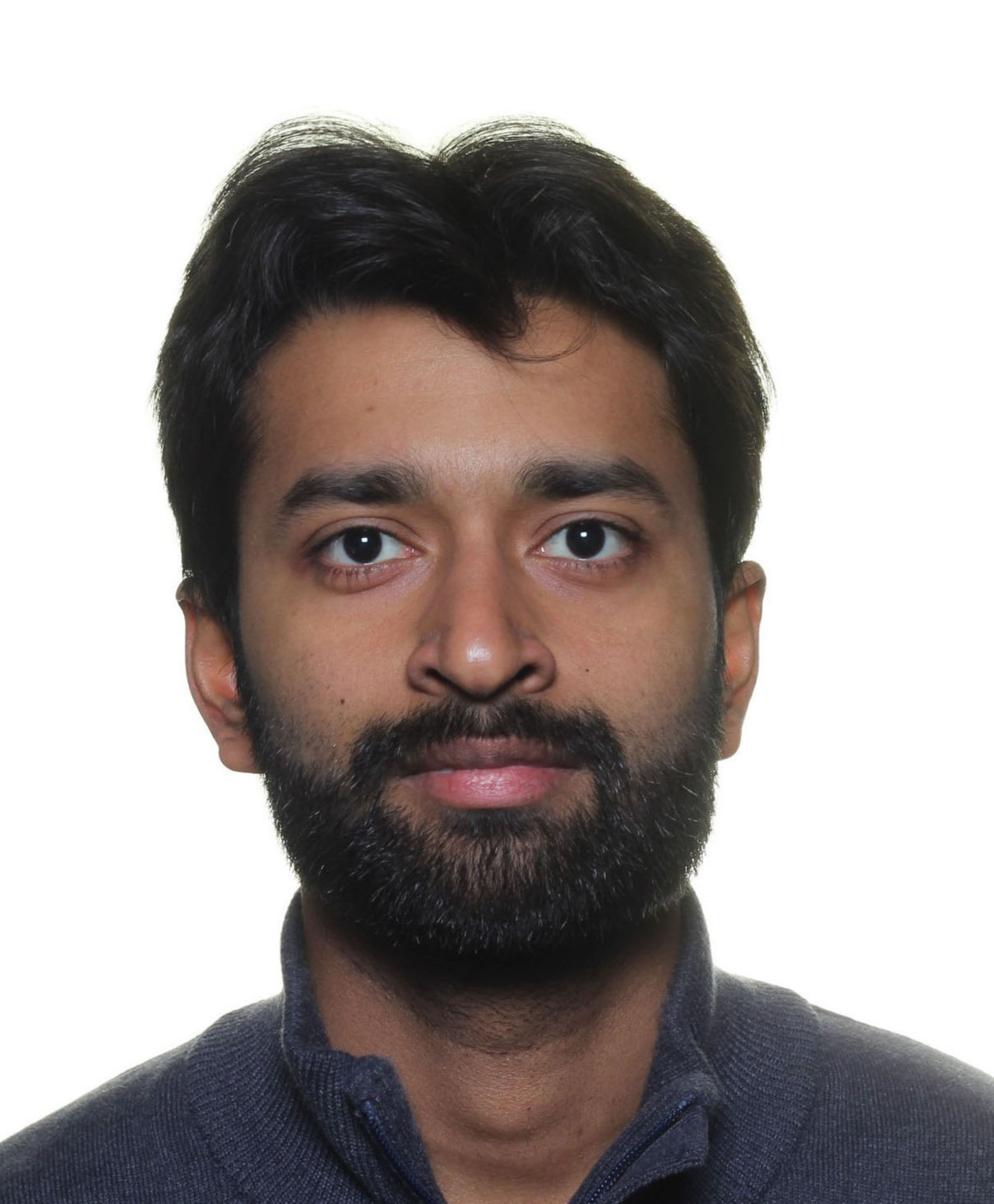}}]{Ayush Bharti} received the B.E. degree in electrical and electronics engineering from Birla Institute of Technology and Sciences, Pilani, India, in 2015, and the M.Sc. degree in signal processing and computing from Aalborg University, Denmark, in 2017, where he is currently pursuing the Ph.D. with the Department of Electronic Systems. His research interests include likelihood-free inference, statistical signal processing, and radio channel modeling.
\end{IEEEbiography}

\begin{IEEEbiography}[{\includegraphics[width=1in,height=1.25in,clip,keepaspectratio]{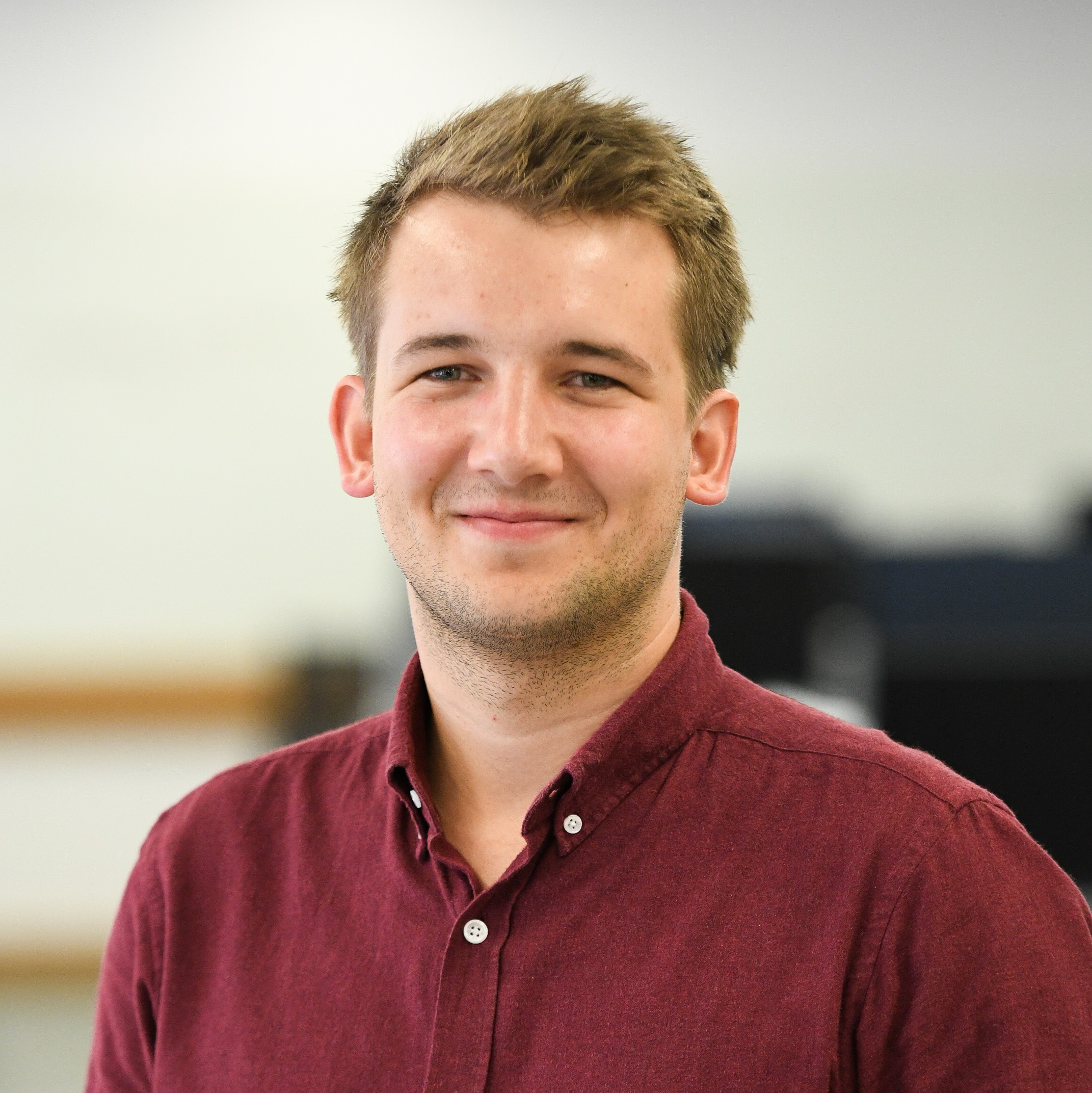}}]{Fran\c{c}ois-Xavier Briol} received a Bachelor with integrated Masters in Mathematics, Operational Research, Statistics and Economics (MMORSE) in 2014, then a PhD in Statistics in 2019, both from the University of Warwick. He was briefly a research associate at Imperial College London and the University of Cambridge, before joining University College London (UCL) in 2019 where he is now a Lecturer in Statistical Science. Dr. Briol is also a Group Leader in Data-Centric Engineering at The Alan Turing Institute, the UK's national institute for Data Science and AI, where he currently leads a project on ``Fundamentals of Statistical Machine Learning''. His research interests include statistical computation and inference for large scale and computationally expensive probabilistic models.
\end{IEEEbiography}

\begin{IEEEbiography}[{\includegraphics[trim={800 300 800 300},clip, scale = 0.75,keepaspectratio]{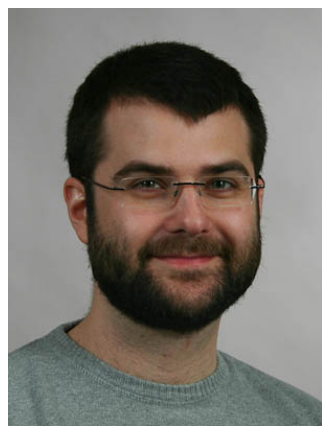}}]{Troels Pedersen} received the M.Sc. degree in digital communications and the Ph.D. degree in wireless communications from Aalborg University, Aalborg, Denmark, in 2004 and 2009, respectively. In 2005, he was a Guest Researcher with the FTW Telecommunications Research Center Vienna, Vienna, Austria. He joined the Department of Electronic Systems, Aalborg University, as an Assistant Professor, in 2009, and became an Associate Professor in 2012. In 2012, he was a Visiting Professor with the Institut d’Électronique et de Télécommunications de Rennes, University of Rennes 1, Rennes, France. His current research interests include statistical signal processing and communication theory, including sensor array signal processing, radio geolocation techniques, radio channel modeling, and radio channel sounding. 
Dr. Pedersen received the Teacher of the Year Award from the Study Board for Electronics and IT, Aalborg University, in 2011 and 2017. 
\end{IEEEbiography}

\end{document}